%
%

\ifx\mnmacrosloaded\undefined \input mn\fi


\pageoffset{-2.5pc}{0pc}

%
  
%
%
%
%
%
\def\teff{{\sl T}_{\rm eff}}
\def\mast{{\sl M}_\ast}

\def\msol{{\sl M}_\odot}
\def\lsol{{\sl L}_\odot}

\def\rast{{\sl R}_\ast}

\def\mast{{\sl M}_\ast}

\def\g{{\it g\/}}
\def\p{{\it p\/}}

\def\bv{Brunt-V\"ais\"al\"a }

\def\sigr{\sigma_{\rm R}}
\def\sigi{\sigma_{\rm I}}
\def\sigc{\sigma_{\rm C}}
\def\sigac{\sigma_{\rm ac}}
\def\kapt{\kappa_{\rm T}}

%
%
\def\unity{ \hbox{1\kern-.23em l} }
\def\field{ \hbox{I\kern-.23em K} }

\def\braket #1.#2.{\langle #1 \vert #2 \rangle}
\def\operator #1.{{\bf #1}}

\def\diff{{\rm d}}

\def\imag{{\rm i}}
\input epsf

\begintopmatter  
\title      { How to drive roAp stars
            }
\author     { Alfred Gautschy$^{1}$,  Hideyuki Saio$^2$,
              Housi Harzenmoser$^{1}$, 
            }
\affiliation{$^1$ Astronomisches Institut der Universit\"at Basel,
              Venusstr.7, CH-4102 Binningen, Switzerland
            }
\affiliation{$^2$ Astronomical Institute, Tohoku University, 
              Sendai, Japan
            }



\abstract { Rapidly oscillating Ap stars constitute a unique class 
            of pulsators to study nonradial oscillations under
            some~--~even for stars~--~unusual physical conditions.
            These stars are chemically peculiar, they have strong
            magnetic fields, and they often pulsate in several
            high-order acoustic modes simultaneously.  We discuss here
            an excitation mechanism for short-period oscillation modes
            based on the classical $\kappa$ mechanism. We particularly
            stress the conditions that must be fulfilled for
            successful driving.  Specifically, we discuss the
            r{\^o}les of the chemical peculiarity and strong magnetic
            field on the oscillation modes and what separates these
            pulsators from $\delta$ Scuti and Am-type stars.  
          }

\keywords { stars; peculiar, Ap~--~stars; oscillations~--~stars; structure of 
          }

\maketitle  

\section{Characterizing the breed}

Rapidly oscillating Ap (roAp) stars constitute a low-temperature subgroup of
the chemically peculiar A-type (Ap) stars with strong Sr, Cr, and Eu
overabundances. The first star of this class, HD~101065, was discovered
photometrically in 1978 by Kurtz (Kurtz 1978, Kurtz \& Wegner 1979).
Presently, the group counts about 30 members.  The periods of the light
variability, which can be multi-periodic, range from roughly 4 to 16 minutes.
As the roAp stars lie in the instability strip close to the main-sequence they
are thought to have masses between 1.5 and $2 \msol$. Therefore, the short
periods must be interpreted as high radial-order \p~modes.  The amplitudes of
variability in the blue, where they are largest, stay below about a hundredth
of a magnitude.

The roAp stars show strong surface magnetic fields exceeding 
frequently some kilo Gauss (cf. Mathys \& Hubig 1997) with
the magnetic axis being inclined relative to the star's rotation
axis. The rotation with periods above 2 days leads, therefore, to a
modulation of the magnetic field strength.  This property gave rise to
the {\it oblique rotator} model (Stibbs 1950). From the temporal phase
behavior of the magnetic field during a rotation period, a dominant
dipole topology was derived for these stars' global magnetic fields.

The observed chemical peculiarities are interpreted as being a very
superficial phenomenon which is restricted to the upper stellar
atmosphere.  It is commonly believed that the strong magnetic field
suppresses turbulent motion which otherwise quickly destroys any
diffusively built up segregation of atomic species that react
differently on the prevailing stellar radiation field (e.g. Michaud
1980, Vauclair \& Vauclair 1982, Alecian 1986). None of the available
diffusion theories reproduces, however, quantitatively the observed
variety of peculiar-star characteristics with a satisfactory low
number of freely tunable parameters.

The short-period oscillation amplitudes of the roAp stars are
temporally not stable but they vary cyclically in phase with the
magnetic field modulation.  The variability is in accordance with the
postulation of dipole oscillation-modes being aligned with the
magnetic axis of the roAp stars which changes its aspect angle during
the star's rotation period.  This picture~--~called the {\it oblique
pulsator} model~--~is furthermore supported by phase jumps of the
pulsational phase at quadrature of the magnetic axis with respect to
the line-of-sight. If the the temporal variation of the pulsation is
parameterized by $\cos (\omega t + \varphi)$, then the phase $\varphi$
jumps by $\pi$ radians at the moment of magnetic polarity reversal;
and this is just what is observed (cf. Kurtz 1990).

The roAp variability is unique in several respects. Despite these
stars sharing their territory in the instability strip with $\delta$
Sct stars, they oscillate with much shorter periods. Much higher
overtones are obviously excited in roAp stars. The mode density in the
corresponding frequency domain is high, and it remains unclear what
selection mechanism is picking out the very low number of observed
oscillation modes. The spherical degrees of the modes identified in
some cases (e.g. Kurtz 1990) were axisymmetric dipole modes. In
general, they must definitely all be axisymmetric, low-$\ell$ modes.

Rotation and magnetic field of the roAp stars cause the $m$-degeneracy
of the oscillation frequencies to be lifted. From the magnitude of the
frequency splitting and the amplitudes of the multiplet components,
aspect angles of the rotation and magnetic axis and spatially
integrated magnetic field strengths can be deduced. If several radial
orders of oscillation modes are observable (as e.g. in HR~1217, Kurtz
et al. 1985) frequency spacings can be deduced. As roAp stars
oscillate in high radial orders and low spherical degree, asymptotic
pulsation theory, together with other (spectroscopic, photometric,
astrometric) observables might constrain global stellar parameters
such as mass and age (cf. Christensen-Dalsgaard 1988, Brown et
al. 1994).

The lifetimes of oscillation modes in roAp stars range from days to
years. For long-lived modes, cyclically (but not periodic) varying
oscillation frequencies on the timescale of months were reported
recently (Kurtz et al. 1994, Kurtz et al. 1997).  The origin of these
shifts is unclear. Speculations, however, tend towards the action of
magnetic activity-cycles as seen on the sun. All in all, the roAp
stars constitute a unique laboratory to test important domains of
stellar pulsation theory and of stellar-hydromagnetics communicated
through oscillations.  For more details we refer to recent reviews
which addressed comprehensively most~--~theoretical as well as
observational~--~aspects of roAp stars: Kurtz (1990), Shibahashi
(1991), Matthews (1997). 

In Section 2 we address the stellar evolution aspect of the roAp stars
and introduce the models with which we worked in our analyses. Section
3 deals with the nonadiabatic oscillation spectra of the roAp stars
and the possible underlying excitation mechanism.  Success and
limitations of our work are critically evaluated and compared with
results in the literature in Sect.~4.  Finally, we devote an Appendix
to the comparison of adiabatic and nonadiabatic oscillation spectra.

\section{Aspects of structure and evolution}
The overlapping of the $\delta$ Sct variables and of roAp stars in the
instability strip on the HR diagram suggests that the latter have
similar masses and that they are also in a comparable evolutionary
stage.  This is additionally supported by the asymptotic frequency
separations (cf. Appendix) which, for $\delta$ Sct-like stars, are
compatible with those seen in roAp stars (e.g. Gabriel et al. 1985,
Shibahashi \& Saio 1985, Heller \& Kawaler 1988).  Therefore, we analyzed
stellar models~--~representing roAp stars~--~with masses between 1.5
and 1.87 $\msol$ during the main-sequence and the early hydrogen 
shell-burning  evolutionary phase (cf. Fig.~1).

\beginfigure{1}

\epsfxsize = 8.3 cm
\epsfbox {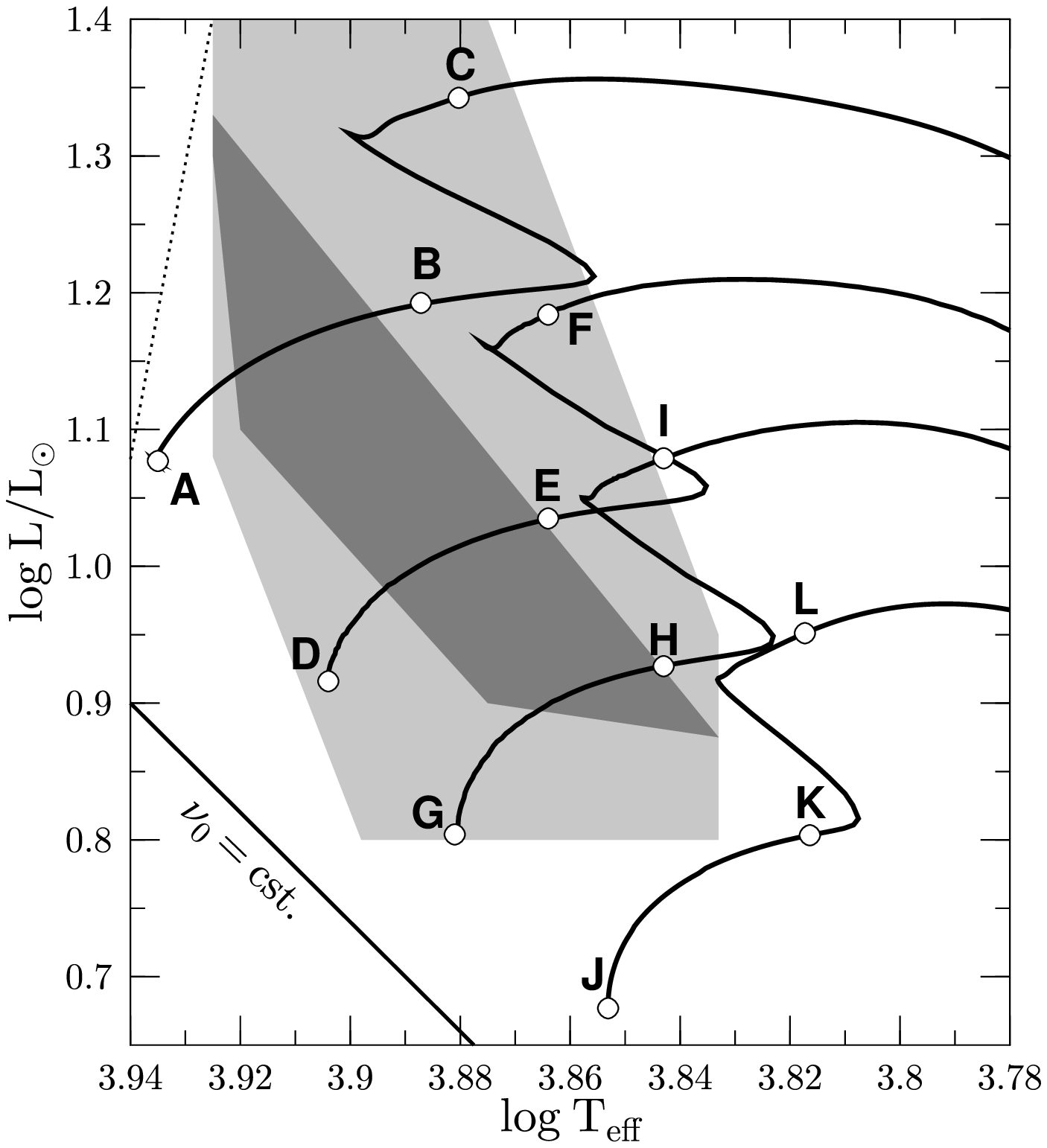}

\caption{{\bf Figure 1.} 
         Hertzsprung-Russell diagram with the loci of the early tracks
         of main-sequence and early post-main sequence evolution of
         stars with 1.5, 1.6, 1.7, and 1.87 $\msol$. The positions along the
         tracks where stellar models were analyzed for their
         oscillation behavior are marked with capital letters. 
         The corresponding parameters are listed in Tab.~1. The grey
         areas denote the map of observed roAp stars onto the HR plane
         as published by North et al.~(1997). For details
         see the text. The dotted line indicates the location of
	 the observed blue edge of $\delta$ Sct stars.
        }
\endfigure

Referring to Hipparcos observations, North et al. (1997) attempted
recently to pin down the location of the roAp stars on the HR diagram.
The result is plotted in Fig.~1 with grey shading. The dark grey area
enclosed the mean values obtained for the sample of 14 stars.  The
lighter grey encloses the maximum extensions of the error bars
specified in North et al. (1997). Their data suggest that stars with
masses above roughly 1.6 $\msol$ could become roAp stars under
suitable conditions mainly during their main-sequence stage.  The
dotted line in Fig.~1 indicates the position of the observed blue edge for
$\delta$ Sct variables. It is evident that an overlap exists;
therefore, roAp and $\delta$ Sct stars appear to share common ground.

For reference, the locus of $\nu_0 = {\rm const.}$ is plotted on the
HR diagram in Fig.~1.  The mathematical definition of $\nu_0$~--~the
asymptotic frequency spacing of equal-degree modes~--~is given in the
Appendix.  The determination of this observable, together with a
reliable temperature measure, constitutes an important
characterization of a star. The $\nu_0$ values of our models are
listed in Tab.~1.

\beginfigure{2}

\epsfxsize = 8.3 cm
\epsfbox {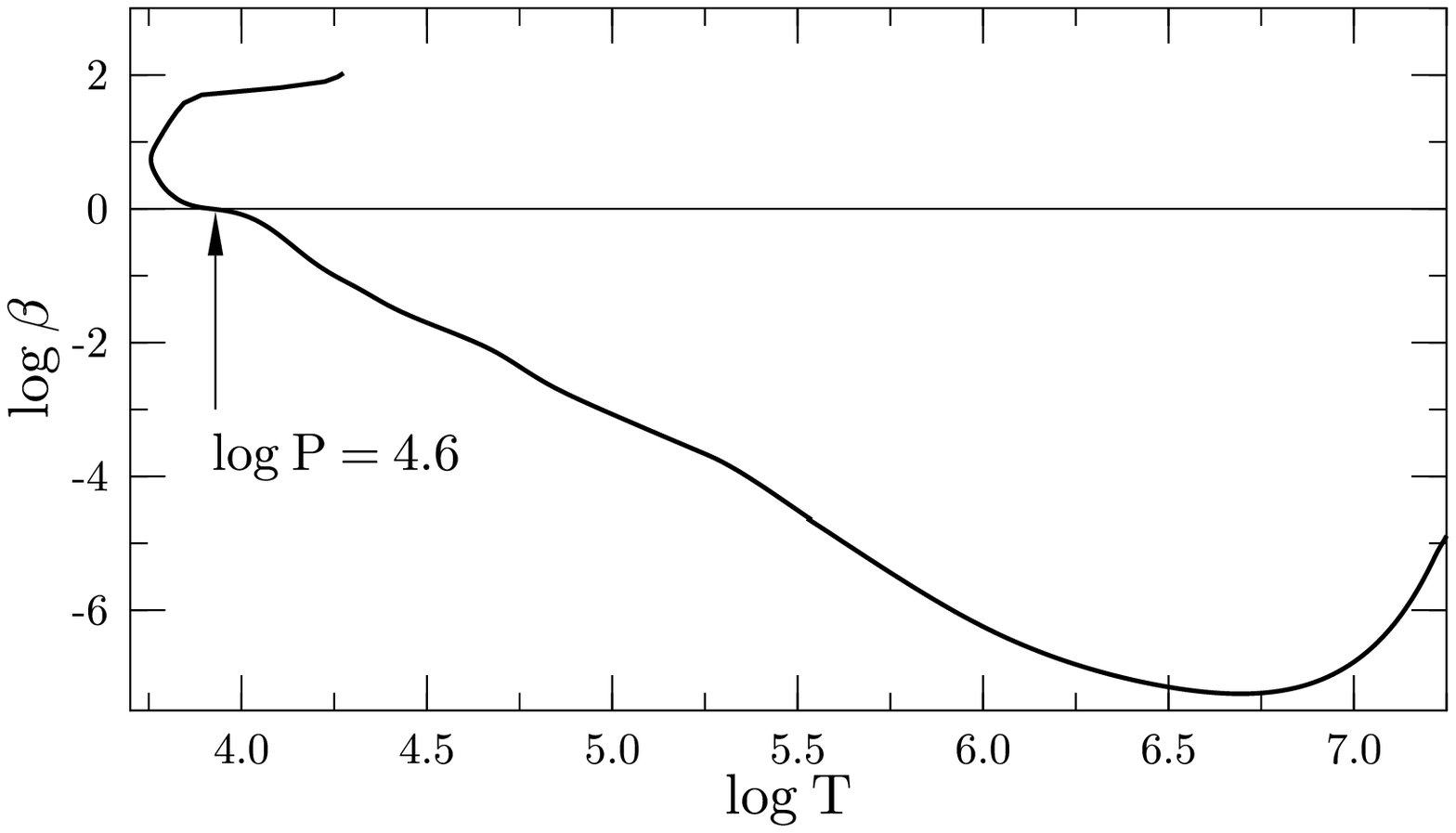}

\caption{{\bf Figure 2.} 
         Logarithm of $\beta$, the ratio of magnetic pressure to total pressure
         throughout a star in the direction of the dipole magnetic pole. 
         The surface magnetic field strength is chosen to be $10^3$ Gauss.
         The stellar model corresponds to model B which is characterized 
         in Tab.~1.
        }
\endfigure

\begintable*{1}
 \caption{{\bf Table 1.} Stellar parameters of models used
          for stability analyses.}
\halign{#\hfil  & \quad\hfil # \hfil\quad &
                  \quad\hfil # \hfil\quad &
                  \quad\hfil # \hfil\quad &
                  \quad\hfil # \hfil\quad &
                  \quad\hfil # \hfil\quad \cr  
         Model  &  $\log \teff$  &  $\log L/\lsol$  &  $M/\msol$  &
  $\nu_0 / \mu{\rm Hz}$  &  $10^7 \times 3 G M / R^3$                 \cr
          A     &     3.936      &   1.086          &   1.87      &
    88.36                &    5.626  
\cr
          B     &     3.884      &   1.240          &   1.87      &
    51.46                &    1.611  
\cr
          C     &     3.880      &   1.352          &   1.87      &
    39.37                &    1.036  
\cr
          D     &     3.904      &   0.916          &   1.70      &
    89.89                &    5.917  
\cr
          E     &     3.864      &   1.035          &   1.70      &
    56.56                &    2.257  
\cr
          F     &     3.864      &   1.170          &   1.70      &
    44.26                &    1.349  
\cr
          G     &     3.880      &   0.804          &   1.60      &
    90.30                &    5.967  
\cr
          H     &     3.843      &   0.927          &   1.60      &
    57.46                &    2.308  
\cr
          I     &     3.843      &   1.079          &   1.60      &
    44.77                &    1.365  
\cr
          J     &     3.853      &   0.686          &   1.50      &
    90.79                &    5.712  
\cr
          K     &     3.816      &   0.812          &   1.50      &
    57.67                &    2.217  
\cr
          L     &     3.816      &   0.960          &   1.50      &
    45.34                &    1.348  
\cr
}
\endtable

The stellar evolution computations for this study were performed with the
Basel stellar evolution code using the micro-physics described in Gautschy et
al. (1996).  The homogeneous chemical composition of the stellar models on the
ZAMS was in all cases $X=0.7, Y=0.28$.  When evolving the stars we did not
account for any chemical peculiarity. We assumed that an atmosphere with a
prescribed $T-\tau$ relation is overlying the stellar photosphere where the
stellar structure computations stopped. Any changes in chemical composition
and molecular weight which characterize Ap-star peculiarity were restricted to
these atmosphere calculations. Descriptions and results of specific choices
are presented in Sect.~3. Furthermore, we did not consider effects of a dipole
magnetic field in the computation of the stellar structure. In this respect it
is worthwhile to notice that even a kilo-Gauss non-potential field is hardly
of relevance for most of the star: Figure~2 shows the logarithm of the ratio
of the magnetic pressure to the total pressure at the magnetic pole (for a
dipole field) for stellar model B. Evidently, the magnetic pressure exceeds
the hydrostatic pressure in the very outermost regions only. Since the the
stellar envelopes of the stellar models corresponding to roAp stars are mainly
radiative they approach quickly the radiative-zero solution.  Therefore,
surface information is not communicated to the deep interior which drives
the star's evolution. The run of $\beta$ in Fig.~2 stops at the outer edge of
the convective central region where the the dipole nature probably breaks
down.

From the stellar evolutionary sequences we selected a few models on
which the stellar oscillation computations were performed.  For each
of the tracks~--~1.5, 1.6, 1.7, and 1.87 $\msol$~--~we selected two
models each in the core hydrogen-burning phase and one in the early
hydrogen shell-burning stage as representatives of roAp stars.  Table 1
lists, for models A to L, some relevant stellar and oscillatory
parameters.

\section{The rapid oscillations of A{\lowercase{p}} stars}
In this section we focus on pulsational aspects of roAp stars.  First,
we address their oscillation-frequency spectra and the problems they
pose in understanding the underlying stellar models.  The later part
of Sect.~3 is devoted to nonadiabatic aspects. In particular we
suggest an excitation mechanism for the observed high-order \p~modes.

The frequently occurring dimensionless frequencies $\sigma =
(\sigr,\sigi)$ are normalized by $\sqrt{3 G \mast / \rast^3}$.  All the
numerical computations of oscillation properties were performed with
codes similar to the one described in Gautschy et al.  (1996).
 
\subsection{Oscillation-mode physics}

To get an impression of the mode cavities in stars it is appropriate to plot
the characteristic frequencies $\sigc$. They separate the spatially
propagative from the evanescent regions in a star (see e.g. Unno et al. 1989).
We compute the dispersion relation of the adiabatic Cowling equations assuming
a plane-wave with a spatial behavior of $\exp(\imag k \cdot x)$ in any
perturbed physical quantity, where we use $x \equiv r / R_\ast$.  Searching
for the loci of Real$(k) = 0$ leads to a quadratic for $\sigc^2$:

$$
  {\sigc^2}_{\pm} ={ - \cal{B} \pm \sqrt{\cal{D}} \over \cal{N}}, 
\eqno(1)
$$
where we define
$$
\eqalign{
	{\cal{A}} &\equiv \left ( V \left ( {2 \over \Gamma_1} - 1 
                                  \right ) - 3
                        \right )^2 ,
	\quad
        {\cal{B}}  \equiv - {A^\ast V \over {\Gamma_1}} 
                          - \ell \left ( \ell + 1 \right ) 
	                  - {{\cal{A}} \over {4} } ,
								   \cr
        {\cal{D}} &\equiv {\cal{B}}^2 
                      - {4\/ \ell \left ( \ell + 1 \right ) A^\ast V
	                   \over
	                    \Gamma_1
                          } ,
	\quad
        {\cal{N}}  \equiv {6 c_1 V \over \Gamma_1} .
                                                                   \cr
}
$$ 
The quantity 
$A^\ast = - V (1 / \Gamma_1 - \diff \log \rho / \diff \log P)$ 
is essentially the Schwarzschild determinant. For the
meaning of the remaining symbols we refer to Unno et al. (1989).

\beginfigure{3}

\epsfxsize = 8.3 cm
\epsfbox{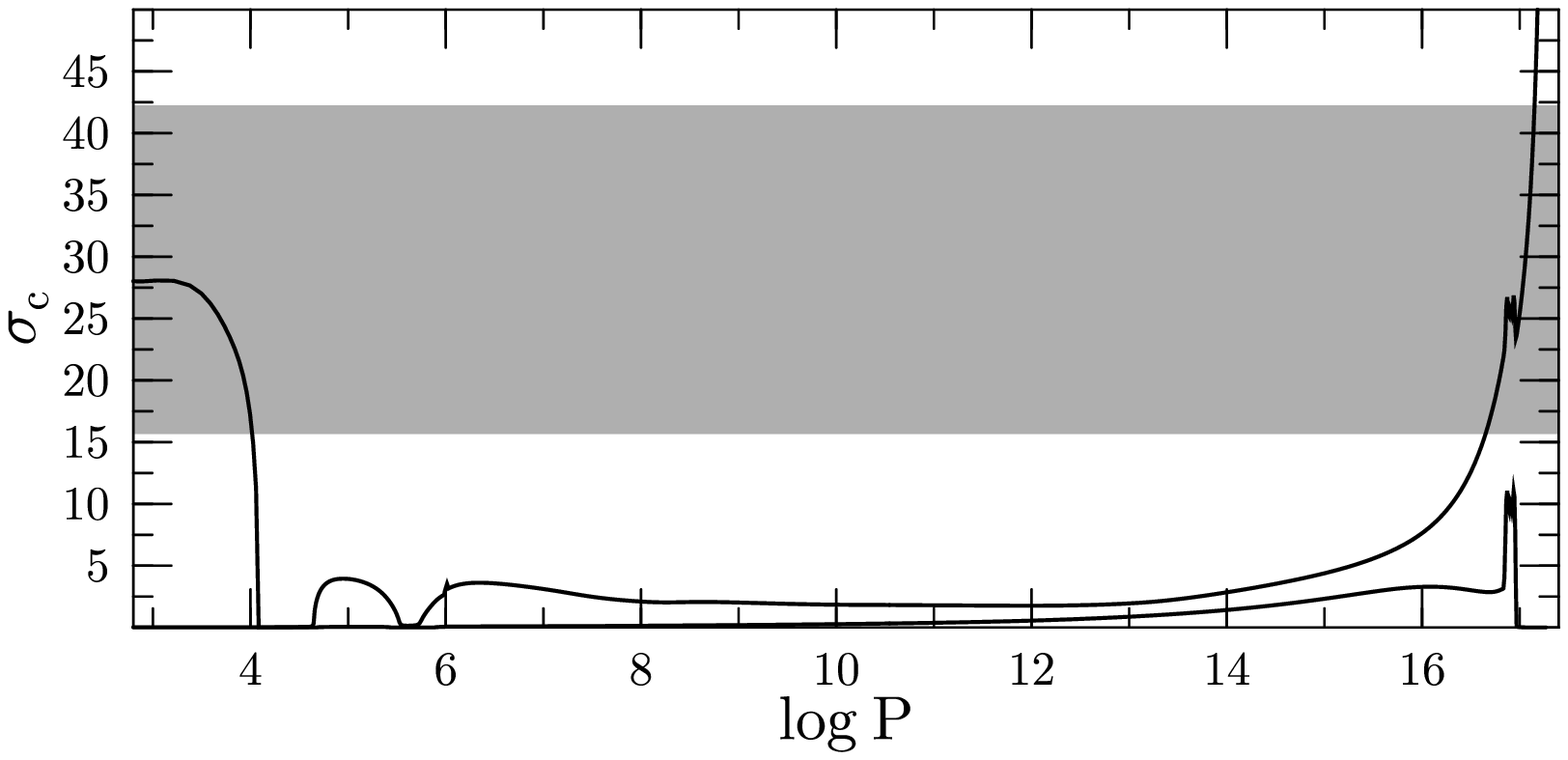}


\caption{{\bf Figure~3.} 
         Characteristic frequencies of dipole modes for model B 
         as function of depth which is measured in $\log P$. 
         The grey-filled area covers the 
         observed oscillation-frequency domain of roAp stars.
        }
\endfigure

As an example of the propagation properties of dipole modes in roAp-like stars
we show in Fig.~3 the characteristic frequencies of stellar model B which is
our slightly evolved $1.87 \msol$ star. The upper and lower lines represent
${\sigc}_+$ and ${\sigc}_-$, respectively. An oscillation mode with frequency
$\sigma$ propagates spatially as a \p~mode in the regions where $\sigma$
exceeds the larger value of $\sigc$.  The dominant features of the upper
$\sigc$ line are the two depressions in the superficial regions which are
caused by partial ionization of H and He ($4 < \log P < 4.7$) and He$^{+}$
($5.5 < \log P < 5.8$). The top of the H/He ionization zone reaches into the
photosphere.  Furthermore, the steep composition gradient which builds up at
the outer edge of the nuclear burning core causes the glitch in $\sigc$ at
about $\log P = 16.8$. The influence on the cavities of the composition change
is even more pronounced in the lower $\sigc$ curve which traces the upper
boundary for the propagation region of the \g~modes.

The oscillation frequency domain observed in roAp stars is marked by
the grey area shown in Fig.~3. We see that all the modes in
the appropriate frequency range are pure acoustic modes.  A slight
influence on the mode spectrum from the steep molecular-weight ($\mu$)
gradient at the outer edge of the nuclear burning convective core
can be expected.

After the roAp variability was interpreted to be due to high-order
acoustic oscillation modes, the magnitude of the critical frequency
($\sigc$) in the stellar atmosphere, i.e. at the outer boundary of the
oscillation cavity, became a point of concern.  The superficial
acoustic cut-off frequency, $\sigac$, which corresponds to $\sigc$
there, determines if a wave with frequency $\sigma$ is reflected (if
$\sigr < \sigac$) or if it is partially transmitted (if $\sigac <
\sigr$) and therefore loses part of its energy into the atmosphere.
Stellar models which are appropriate for roAp stars show the
convection zone induced by H/He ionization to reach up to the stellar
photosphere. At the photosphere the acoustic wave propagation is
determined by the run of the \bv frequency which is imaginary in the
convectively unstable layers. To find the outer edge of the acoustic
cavity, {\it some} atmospheric modeling is necessary. In other words,
the action of the outer boundary condition is dominated by the
peculiarities in the optically thin radiative layers. The use of
standard (e.g. Eddington-grey) $T - \tau$ relations leads
to rather low cut-off frequencies. Many of the observed oscillation
frequencies lie above the cut-off so that some mode energy
is lost from the star (cf. Fig.~3). If the cut-off frequency in roAp
stars were indeed that low, then the excitation mechanism would have
to be very efficient to over-compensate the modal leak into the
atmosphere.

To resolve the dilemma of the low cut-off frequencies at the surfaces,
considerable efforts went into hypothesizing physical processes which
act in the stellar atmospheres to sufficiently increase the acoustic
cut-off frequency. Shibahashi \& Saio (1985) computed critical
frequencies for Eddington-type atmospheres as well as for $T - \tau$
relations fitted to more detailed Kurucz atmospheres. Their fitted
solutions showed a larger ratio of $\teff/T(\tau=0)$ than in the
Eddington atmospheres. For none of the chosen $T - \tau$ functions,
however, a $\sigac$ resulted that would lead to fully reflected
roAp-relevant oscillation modes.  The results of an observational
determination of the $T - \tau$ relation of a roAp star were presented
by Matthews et~al.~(1996). Their studied star, HR~3831, indicated an
even steeper temperature drop close to the photosphere and a more
isothermal behavior at small optical depths than what Shibahashi \&
Saio (1985) suggested. (It is, however, noteworthy that Kurtz \&
Medupe (1996) in their analysis of frequency-dependent oscillation
amplitudes conclude that a $T-\tau$ relationship cannot be derived
from roAp oscillation data.)  A list of theoretical suggestions to
increase $\sigac$ at the outer boundary of the stellar models was
given in Section~5 of Shibahashi (1991).

\beginfigure{4}

\epsfxsize = 8.3 cm
\epsfbox{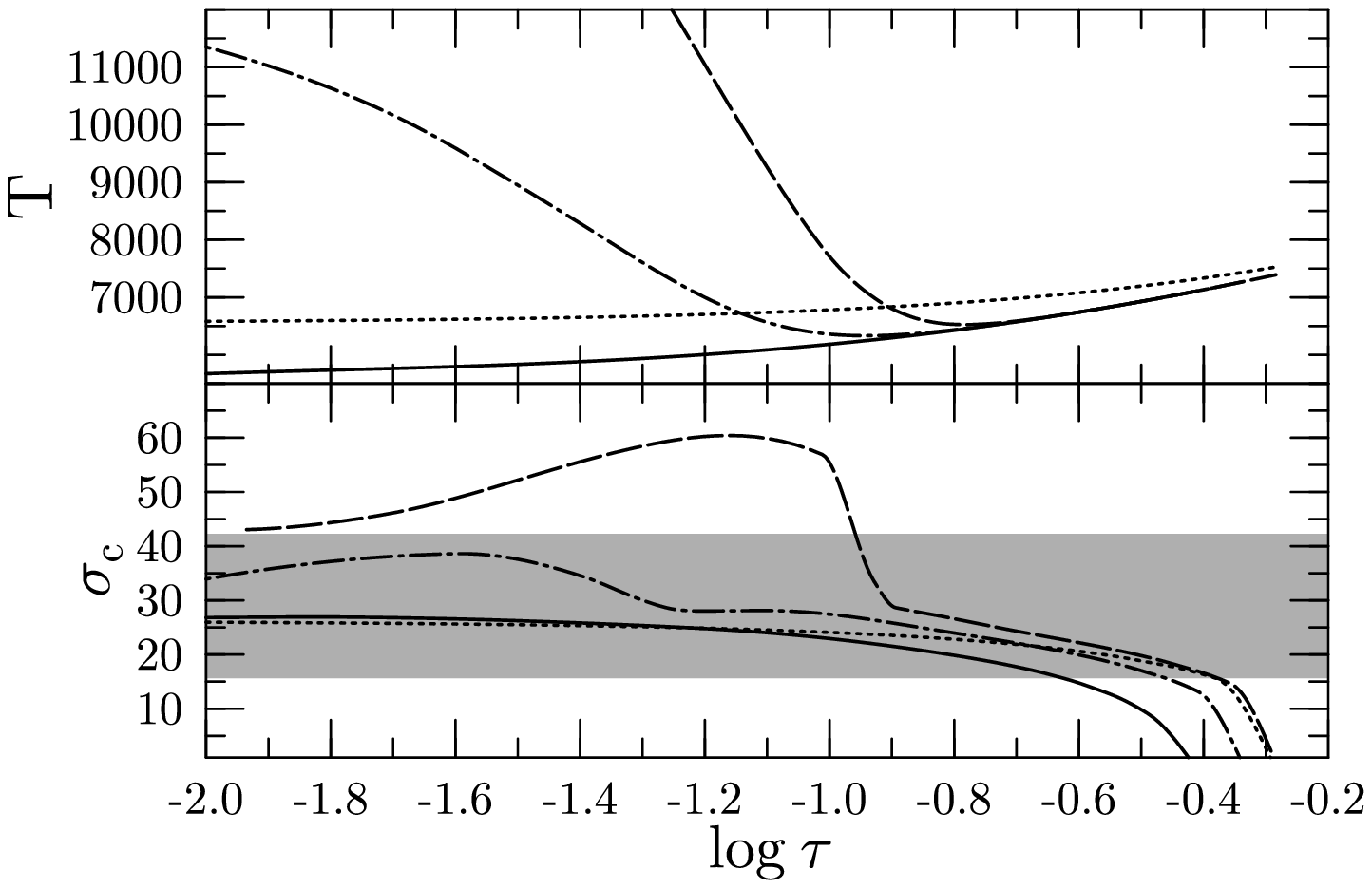}


\caption{{\bf Figure~4.} 
         Dependence of the characteristic frequency $\sigc$ (bottom
         panel) in the superficial layers  
         on the choice of the $T - \tau$ relation (top panel) 
         in the star's atmosphere. The case of dipole modes
         in model B are depicted. 
        }
\endfigure

To simulate the effect of roAp-type chemical inhomogeneities on the
characteristic frequencies, in particular on the atmospheric cut-off
frequency, we studied sharp $\mu$ transitions in atmospheres with standard $T
- \tau$ relations.  According to eq.~3, the floating on the top of the stellar
atmosphere of rare earth elements leads to a $\mu$-{\it inversion\/}.  The
magnitude and the location of such $\mu$-jumps were guided by results of
Michaud~(1980). Our atmospheric $\mu$ variations led to $\sigc$ depressions
which had no noticeable influence on the amount of energy leakage of the
waves with frequencies above the cut-off frequency.  Acoustically, the feature
was found to be too shallow to be noticed even by high-frequency \p~modes.

We postulate now a mechanism which rather successfully increased $\sigc$ in
our models. It remains to be seen, however, if it is indeed realized in
nature: Since roAp stars have a convection zone which reaches the photosphere
and since they have strong magnetic fields, we suggest that roAp stars are
capable of building up ~--~very much like solar-type stars~--~a chromosphere
of some extent. This chromosphere goes along with a temperature inversion
which we ad hoc parameterized by:
$$
\eqalign{  
           T(\tau) & = 0.931 \cdot\teff\cdot
             [      \tau   + 0.72858 
                            - 0.3367\cdot\exp(-2.54 \tau)       \cr
                   & + 10\cdot\exp(-50 \tau)
                            - 1.0\cdot\exp(-500 \tau)
             ] ^{1/4}.                                        \cr 
        } 
\eqno(2)
$$ 

The above choice is motivated by the Shibahashi \& Saio (1985) fit
(full line in Fig.~4) except that the first coefficient in the square
bracket is adapted to enforce $T(\tau = 2/3) = \teff$. The positive
sign in front of the $\exp(-50 \tau)$ coefficient of eq.~2 induces a
temperature inversion at small optical depths (dash-dotted line in
Fig.~4). Replacing this term by $+120 \cdot \exp(-40 \tau)$ leads to
the stronger temperature inversion delineated with the dashed line.
Finally, the dotted line in Fig.~4 indicates the run of an
Eddington-type $T - \tau$ relation.

We can understand the effect of the temperature inversion on the
{\it p\/}-mode cavity by approximating $\sigc$ for acoustic modes close to 
the surface:
$$
\eqalignno{
 \sigc \approx & {V \over 3} 
               \left ( 
                      {\diff\log \rho \over \diff\log P}
                     - 
                      {1 \over \Gamma_1}
               \right )
                                                       &(3) \cr
               &  =
              {V \over 3}
               \left (
                       \alpha 
                     - \delta \nabla_{\rm T}
                     + {\varphi \over \delta} \nabla_\mu
                     - {1 \over \Gamma_1}
               \right ).                               &    \cr
         }
$$
We used $\alpha = (\partial \log \rho / \partial \log P)_{\rm T}$, 
$\delta = -(\partial \log \rho / \partial \log T)_{\rm P}$, and
$\nabla_{\rm T}\equiv \diff \log T / \diff \log P$ to be the
effective temperature gradient.  A negative $\nabla_{\rm T}$ clearly
increases the critical frequency for adiabatic \p~modes. Quantitative
effects are displayed in Fig.~4.

The temperature inversions shown in Fig.4 are, admittedly, rather strong. In
particular, we have to face the fact that no observational signature for
chromospheric activity has been found to date (e.g. Shore et al. 1987).
Therefore, the conjectured temperature inversion cannot be very strong because
it should have been observable through emission line features otherwise.
Notice, however, that only the innermost part of the above temperature
inversions are essential for the increase of $\sigc$. Hence, a leveling off of
the temperature after an increase of 3\,000 Kelvin at most is sufficient to
completely reflect the roAp-type acoustic waves in many cases.

In the next section we investigate the effect of this modified outer
edge of the cavity on the stability properties of high-order \p~modes.
As of now, the physics of their excitation remains a matter of debate.

\subsection{Excitation physics}

Since the roAp star domain on the HR diagram appears to overlap with
the $\delta$ Sct stars (see Fig.~1), the partial He$^+$ ionization
zone was thought for a long time to be also the driving agent for roAp
stars (e.g. Kurtz 1990, Matthews 1997).  Numerical computations never
supported this conjecture, however. Based on the compositional
peculiarity of the roAp stars, Matthews (1988) hypothesized, relying
on qualitative micro-physical arguments, that Si~IV partial ionization
might provide sufficient {\it p}-mode driving even if He was drained
out of the potential driving depth in the star. He estimated an
overabundance of Si IV in the superficial stellar regions of the order
of 200 to be sufficient for the mechanism to work.  A completely
different avenue was taken by Shibahashi (1983) which was also
investigated by Cox (1984). Their local analyses of oscillatory
magneto-gravity waves suggested overstability under favorable
circumstances. They expected the magnetic polar regions to become
overstable under Ap-star circumstances. The eventually observable roAp
star oscillation are those global \p~modes which are resonantly
excited by the overstable magneto-gravity waves. This picture has the
great advantage of naturally explaining the alignment of the pulsation
axis with the magnetic axis and the selection of axisymmetric modes.
Detailed global analyses which are necessary to settle the excitation
problem eventually are, however, lacking also for this approach.
Recently, Dziembowski \& Goode (1996) noted that the partial H
ionization contributes positively to the work integral. No \p~mode
with a roAp-star relevant period was found yet destabilized by
the $\kappa$ mechanism.

\beginfigure{5}

\epsfxsize = 8.3 cm
\epsfbox{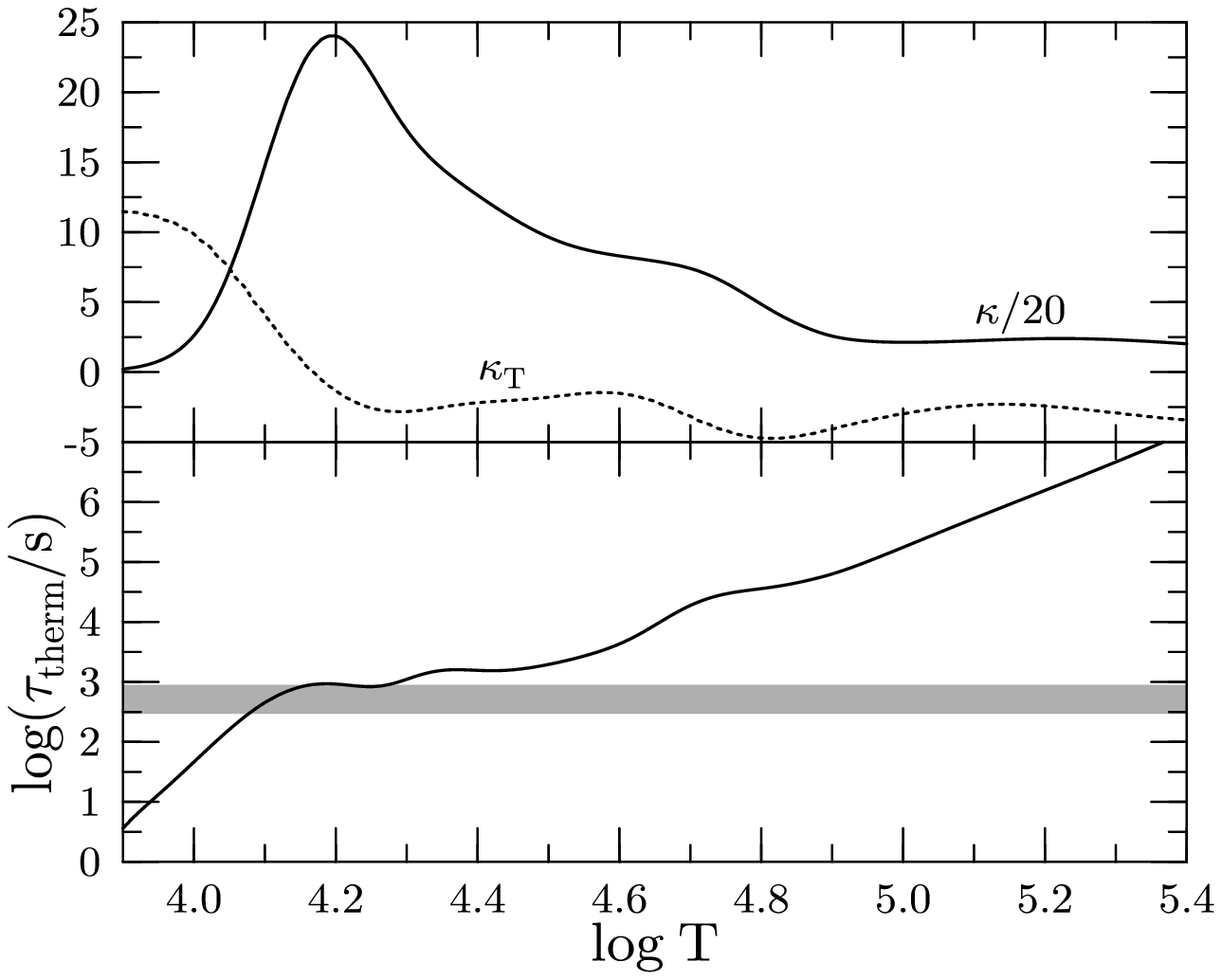}

\caption{{\bf Figure 5.} 
         Spatial run of $\kappa_{\rm T}$, a scaled $\kappa$,
         and (in the lower panel) the variation of the
         thermal time-scale ($\tau_{\rm therm}$) in model J. 
         The grey shading indicates the temporal region 
         of interest for roAp oscillations (cf. text).
        }
\endfigure

If a ``thermo-mechanical valve mechanism'' is to power roAp-like
oscillation modes, then the oscillation periods have to be of the order
of the thermal timescale ($\tau_{\rm therm}$) of the envelope overlying
the driving region (e.g. Cox 1980).  The corresponding spatial
structure shown in Fig.~5 for stellar model~E is representative for
the whole model series looked at in this paper.  A qualitative
inspection of the models shows that those superficial regions
with a thermal timescale below about $1\,000$~s contain the steep
$\kappa_{\rm T}$ rise towards the surface induced by partial H and He
ionization (see Fig.~5). The deeper-lying, potentially driving region
for which $\diff\, \kapt / \diff\, r > 0$ (at $4.1 \la \log T \la 4.8$
in Fig.~5) favors time-scales~--~as expected~--~of modes in the
$\delta$ Sct period domain, i.e. above $10^4$~s. Therefore, we
expect~--~already a priori~--~partial ionization of H and He to be the
prospective excitation agents for roAp-type oscillations.

\beginfigure{6}

\epsfxsize = 8.3 cm
\epsfbox{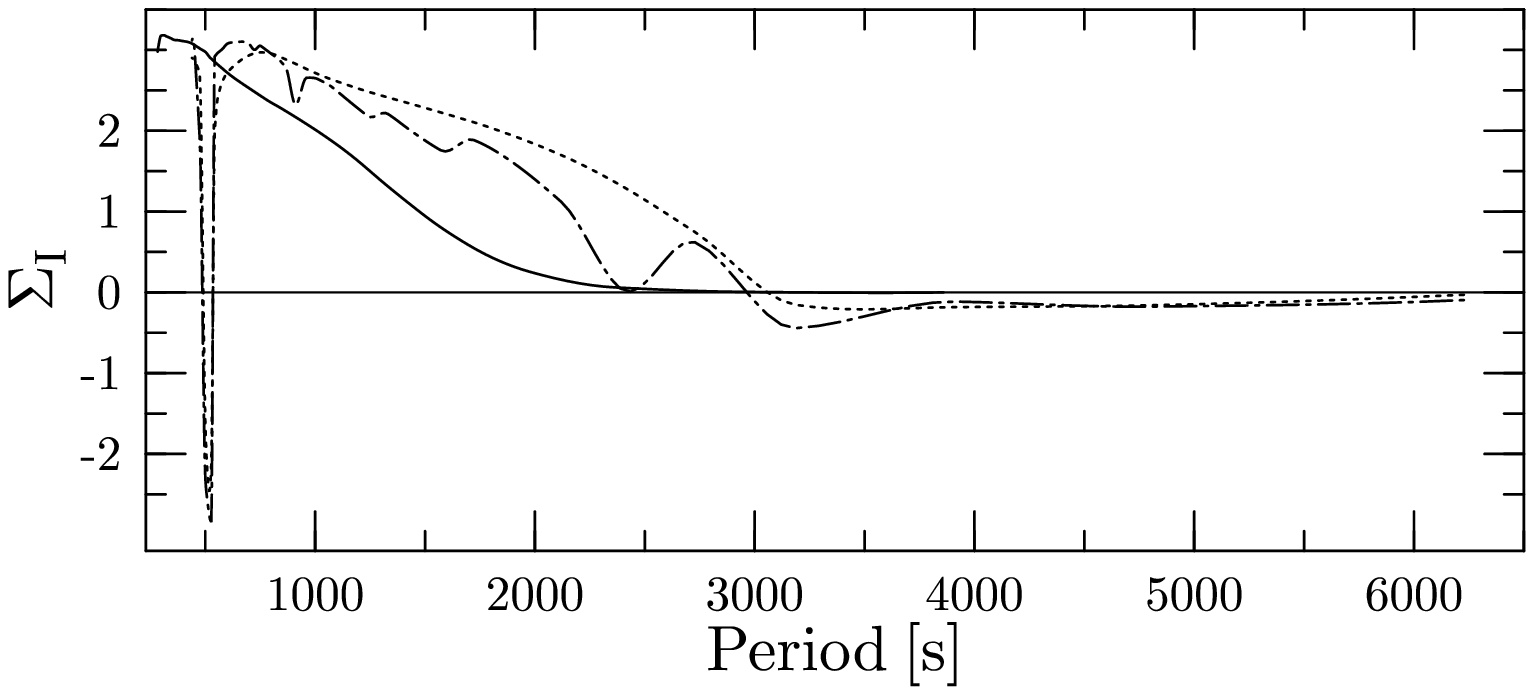}

\caption{{\bf Figure 6.} 
         Appropriately scaled imaginary parts of \p~modes for
         the 3 models of the $1.6 \msol$ sequence with the
         weak temperature inversion. Full line: model G, dotted line:
         model H, and dashed-dotted line: model I.
        }
\endfigure

We computed nonadiabatic oscillation spectra for the stellar models
listed in Tab.~1.  The chromosphere-like temperature structure in the
atmosphere which we introduced provided a sufficiently high acoustic
cut-off so that the acoustic cavity appears reflective at the surface
for most of the computed oscillation modes. The numbers of overstable
\p~modes in the different models are listed in Tab.~2. For the
$\delta$ Sct-type oscillation modes we indicate only {\it if\/} we found any
(``x'' in Tab.~2) or none (``-'' in Tab.~2).  Selected, results are
discussed in the following to demonstrate their properties and the
physical mechanisms.

\begintable*{2}
 \caption{{\bf Table 2.} Excited dipole modes of roAp and $\delta$ Sct type 
           for weak (first entry) and strong (second entry)
           T-inversions and homogeneous envelope compositions.
         }
\halign{#\hfil  & \quad\hfil # \hfil\quad &
                  \quad\hfil # \hfil\quad &
                  \quad\hfil # \hfil\quad &
                  \quad\hfil # \hfil\quad &
                  \quad\hfil # \hfil\quad &
                  \quad\hfil # \hfil\quad &
                  \quad\hfil # \hfil\quad &
                  \quad\hfil # \hfil\quad &
                  \quad\hfil # \hfil\quad &
                  \quad\hfil # \hfil\quad &
                  \quad\hfil # \hfil\quad &
                  \quad\hfil # \hfil\quad \cr  
  Period domain &  A  &  B  &  C  &  D  &  E  &  F  &  G  &  H  &  I
 &  J  &  K  &  L  \cr
  roAp          & 0,0 & 0,0 & 0,0 & 0,0 & 0,0 & 0,0 & 0,0 & 3,3 & 3,2
 & 1,3 & 5,4 & 4,3 
\cr
  $\delta$ Sct  & -,- & x,x & x,- & x,x & x,x & -,- & x,- & x,- & x,-
 & x,x & x,x & x,x
\cr
}
\endtable

Figure~6 shows the whole dipole {\it p}-mode domain in which overstable
oscillation modes were found for the $1.6 \msol$ models.  The
imaginary parts of the eigenfrequencies were suitably scaled to
accommodate the large numerical range of $\sigi$ in the short- and
long-period domain. We defined:
$$
\Sigma_{\rm I} \equiv {\rm Sign} \left ( \sigma_{\rm I} \right ) 
             \cdot \log \left  ( 
             1 + { \vert \sigma_{\rm I} \vert \over 1 \cdot 10^{-4} } 
                        \right ).
\eqno(4)
$$
Negative $\Sigma_{\rm I}$ values represent overstable modes. Values of
$\Sigma_{\rm I}$ exceeding unity behave essentially logarithmic whereas
smaller values are linear in $\sigi$. 

The full line in Fig.~6 (as well as in other figures showing the three
evolutionary stages of a selected stellar mass) stands for the
zero-age main-sequence (ZAMS) model, the dotted line represents the
slightly evolved model close to the first turn-around, and the
dashed-dotted line depicts the results of the most evolved stage. For
all models in Fig.~6 we used the weak temperature inversion.  The $1.6
\msol$ ZAMS model is pulsationally stable over the whole frequency
domain (i.e. roAp and $\delta$ Sct modes). The more evolved models
show a short-period instability around 500 s (the number of
overstable modes is listed in Tab.~2) as well as overstable long-period modes
above 3\,600 s, which are appropriate for $\delta$ Sct variables.  For
both stellar models (H and I) the range of overstable roAp-type modes
remains unchanged. This is different in the case of a strong
T-inversion in the $1.5 \msol$ sequence which is shown in Fig.~7.
In this case the instability domain shifts to longer periods as
the star evolves. On the ZAMS, periods around 400 s are excited
where the subgiant model has 4 overstable \p~modes around 750 s.

\beginfigure{7}

\epsfxsize = 8.3 cm
\epsfbox{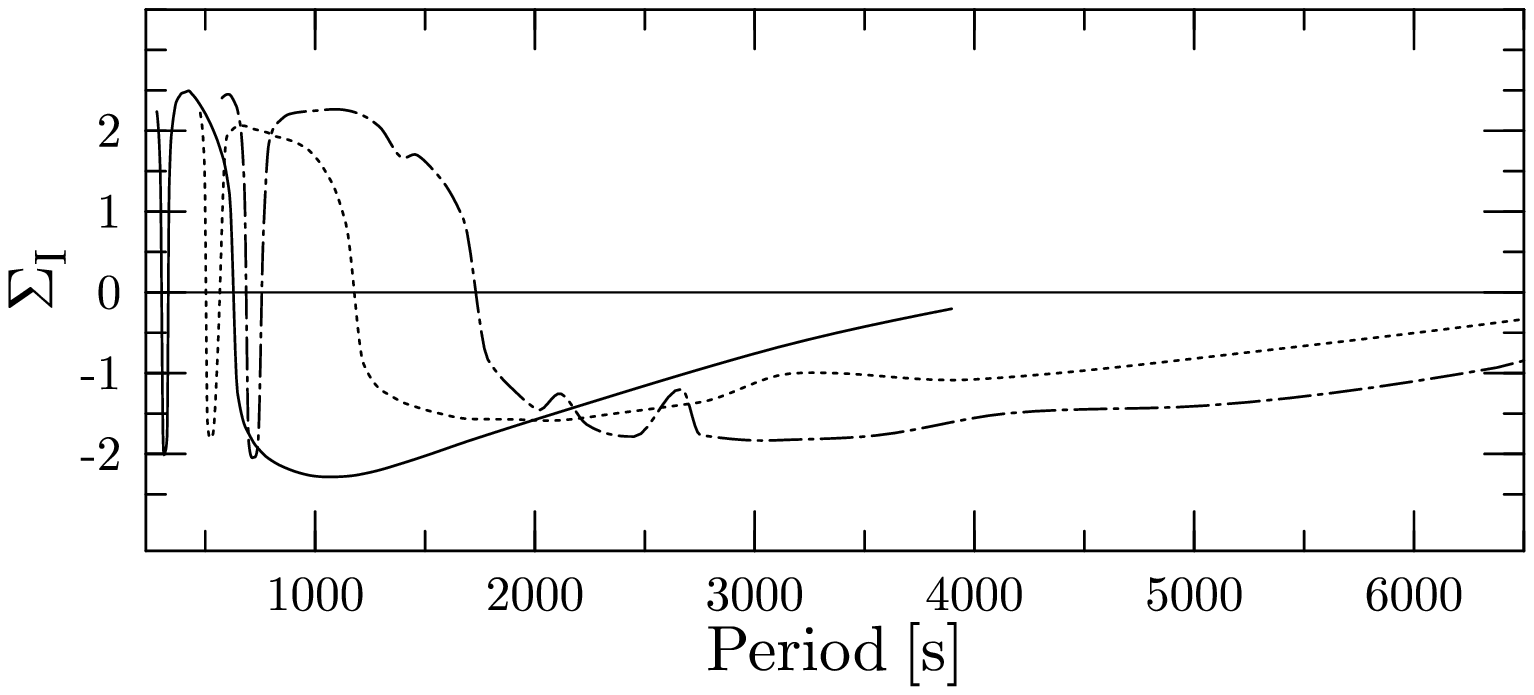}

\caption{{\bf Figure 7.} 
         Appropriately scaled imaginary parts of \p~modes for
         the 3 models of the $1.5 \msol$ sequence and a strong
         temperature inversion. Full line: model J, dotted line:
         model K, and dashed-dotted line: model L.
        }
\endfigure

In all models with overstable high-order \p~modes we found only
a few of them (cf. Tab.~2) at a selected spherical degree. This
short period instability domain is always separated from the long-period
domain with low-order \p~modes which are typical for  $\delta$
Sct variability. It is only in the $1.5 \msol$ ZAMS model where even
this longer-period instability range extends to below 1\,000 s.  

\beginfigure{8}

\epsfxsize = 8.3 cm
\epsfbox{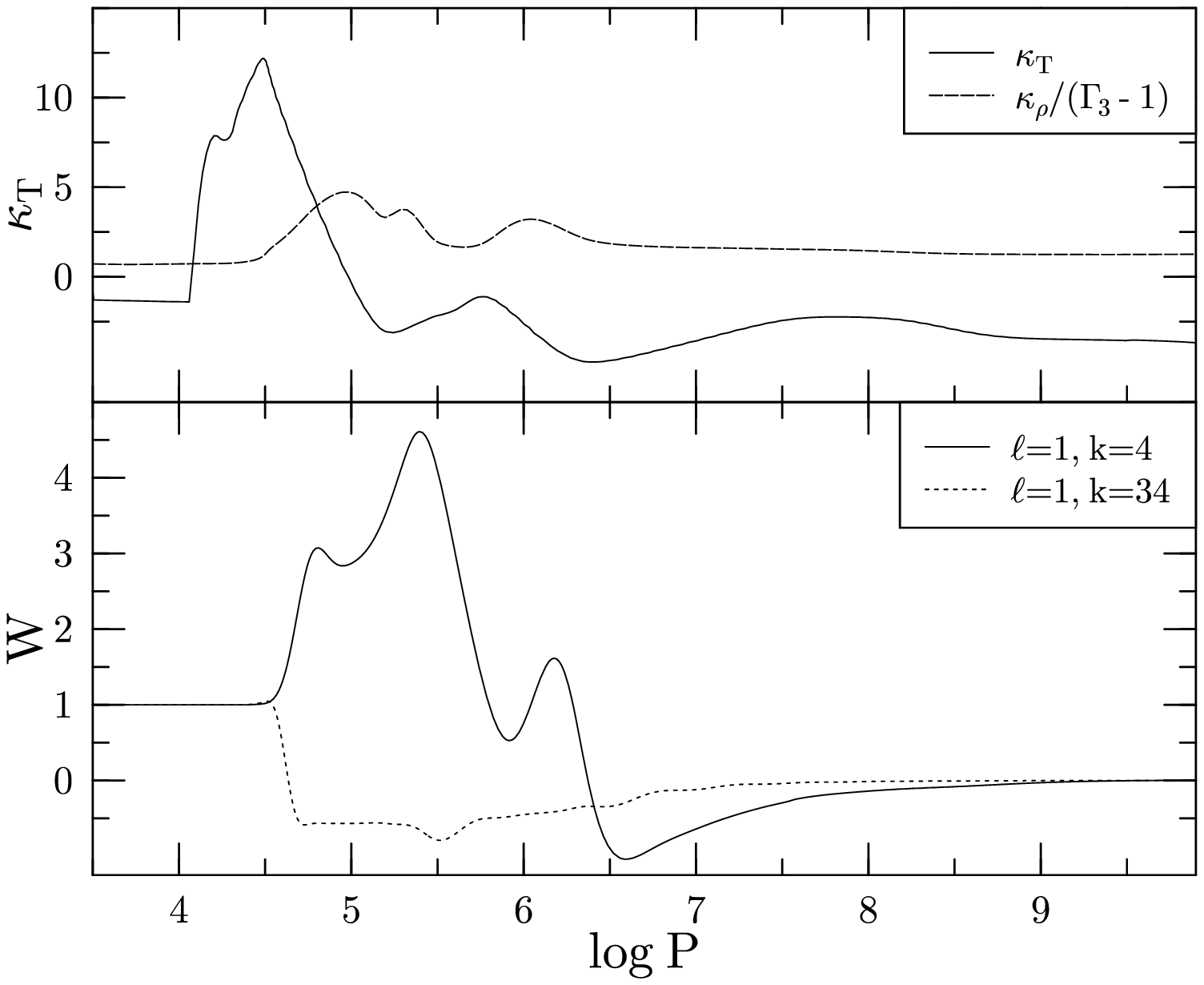}

\caption{{\bf Figure 8.} 
         Top panel shows the spatial run of the opacity derivatives
         which enter the computation of the work integral. 
         The lower panel displays the cumulative work done
         by a short-period (dashed line) and a long-period (full line)
         mode, both of which are overstable.
        }
\endfigure

To study the destabilisation of the \p~modes we plotted some relevant
physical quantities of model H as a function of depth in the envelope
(parameterized by the total pressure) in Fig.~8. The top panel shows
the opacity derivatives along the chosen thermodynamic basis. It
was argued in Unno et al. (1989) that a necessary condition for
pulsational driving requires 
$$
    {\diff \over \diff r}\left ( \kappa_{\rm T} 
                               + {\kappa_{\rho} \over {\Gamma_3 - 1}}
                        \right ) > 0.
\eqno(5)
$$
The major contributor to the above sum is clearly $\kappa_{\rm T}$.
The outer potential driving region is due to partial H and He
ionization. The second, deeper lying one is associated with the He$^+$
partial ionization. We see that the long-period mode is mainly driven
by He$^+$ ionization, the outer ionization zone is actually damping in
this case. The high-order ($k=34$) dipole mode, on the other hand gets
all the driving from the H/He ionization zone. The deeper lying
regions are all slightly damping. The results displayed in Fig.~8
agree very well with the rather qualitative reasoning on the potential
driving agent for roAp modes used before. Notice, however, that
the condition in Eq.~5 was derived for weakly non-adiabatic
circumstances. Therefore, noticeable deviations from this simple
form must be expected  occasionally.

\subsection{Inhomogeneous envelopes}

A persistent problem with the classical $\kappa$ instability mechanism
found above is that not only short-period modes of high radial order
are excited, but also the $\delta$ Sct-type low-orders. Observations
clearly show that $\delta$ Sct-like oscillations are not present in
roAp stars.

To suppress the $\delta$ Sct-type modes we assumed a drainage of He
from the superficial regions.  Stabilisation of turbulent bulk
motion by the magnetic field might cause such a He depletion in the
outer parts of Ap-star  envelopes (Vauclair \& Vauclair 1982).
We assumed ad hoc that the He drainage extends over those layers where
magnetic pressure (for a kilo-Gauss field) exceeds gas pressure.  We
depleted helium to a mass fraction of 0.1 (and enhanced H to 0.88) at
temperatures below $\log T = 4.5$ (cf. Fig.~2). Across a transition
region of $\Delta \log T = 0.1$ our canonical population I composition
of $X=0.7, Y=0.28$ was recovered. For ease of computation, the smooth
transition was prescribed by a hyperbolic tangent function.  Fig.~2
indicated already that only an extremely thin mass layer is affected
by the drainage. In models H, K, and L for which we performed this
experiment the He drainage extended over the outer $4\cdot 10^{-9}
\mast$. Notice that this is six orders of magnitude less than what
Shibahashi \& Saio (1985) adopted in their study.

\beginfigure{9}

\epsfxsize = 8.3 cm
\epsfbox{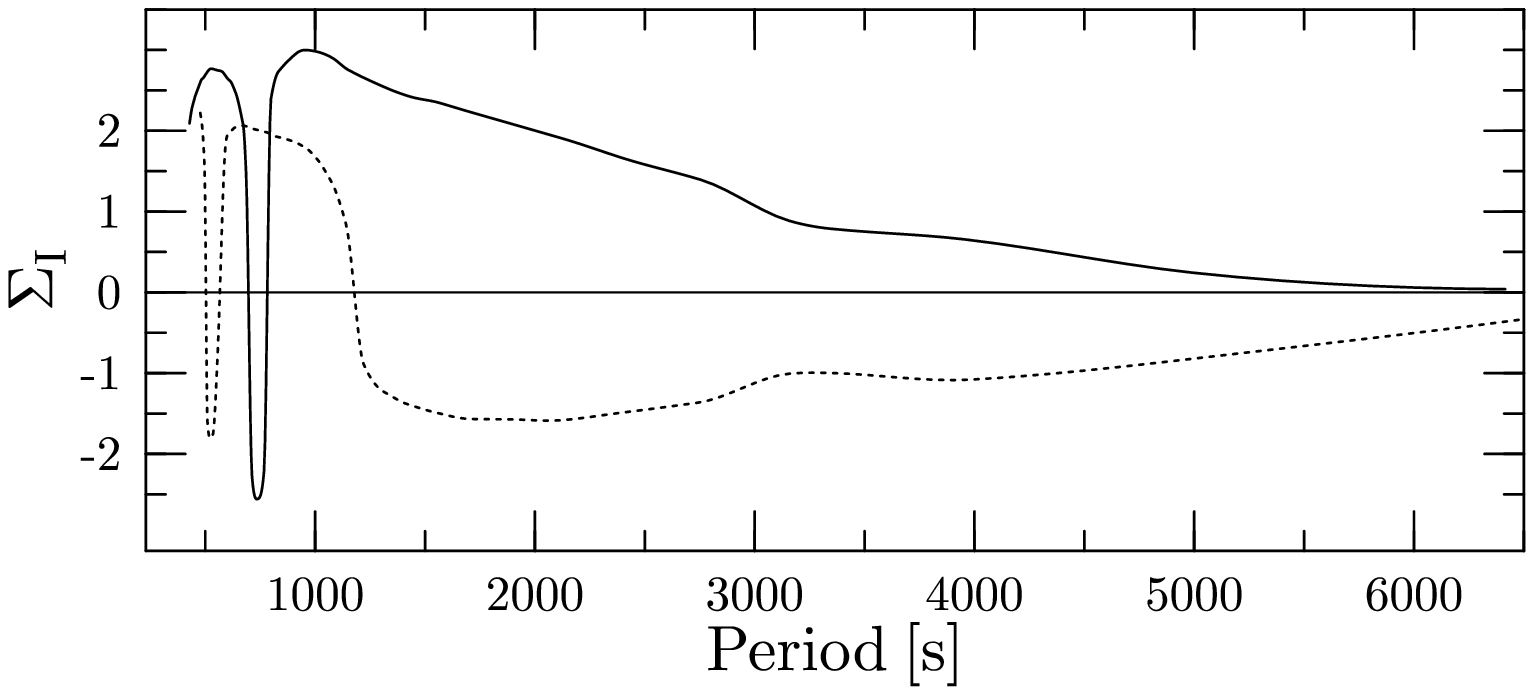}

\caption{{\bf Figure 9.} 
         The run of $\Sigma_{\rm I}$ as a function of period for model
         K with strong temperature inversion. The full line shows the
         results for the inhomogeneous envelope. The dashed line
         depicts the behavior of oscillation modes in the model with
         homogeneous outer layers. 
}
\endfigure

The repeated nonadiabatic stability calculations on models H, K, and L showed
that the long-period instabilities disappear if the superficial layers are
chemically inhomogeneous.  The effect is less pronounced in model H than in
the two $1.5 \msol$ ones, however. Additionally, the oscillation modes of the
strong temperature-inversion models seem to react more strongly to
inhomogeneity than those with the weak inversions. The dipole-mode results of
model K are shown in Fig.~9. The long-period modes above about 1\,200 s which
are overstable in the model with chemically homogeneous envelope (broken line)
are all stabilized in the He-drained one (full line). The short-period modes
remain overstable. The radial orders of the overstable modes are, however,
lower in the inhomogeneous envelope: In the homogeneous outer layers, periods
around 500~s are excited whereas periods of about 750~s are favored in the
inhomogeneous case. This property is not generic, however.  In other cases,
such as model L, the short-period instability domain remained unchanged. A
peculiarity of this model L was that in addition to the short-period
instability a dipole-mode at about 900~s turned overstable in the
inhomogeneous envelope model. In other words, we had one case where {\it
non-successive\/} radial orders were excited. We show the corresponding
diagram in Fig.~10.

\beginfigure{10}

\epsfxsize = 8.3 cm
\epsfbox{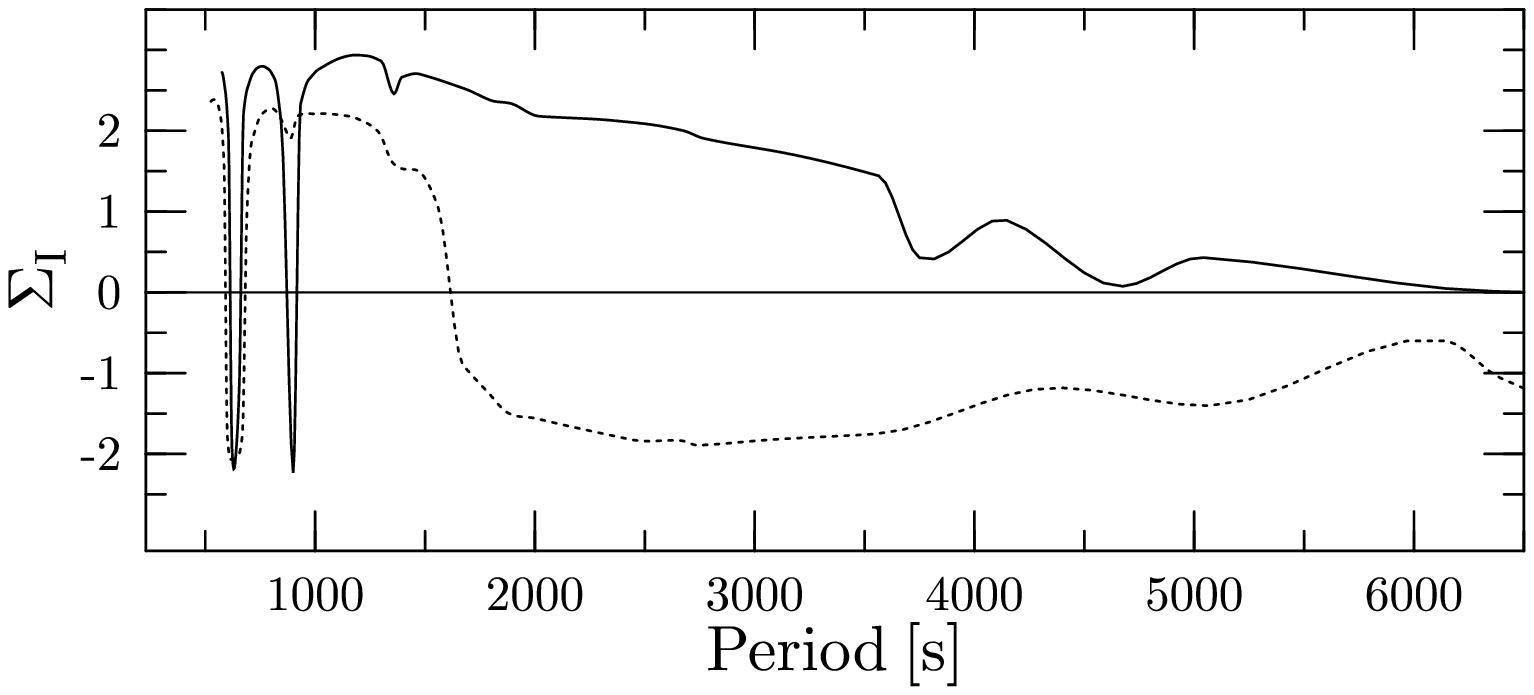}

\caption{{\bf Figure 10.} 
         The run of $\Sigma_{\rm I}$ as a function of period for model
         L with weak temperature inversion. The full line shows the
         results for the inhomogeneous envelope. The dashed line
         depicts the behavior of oscillation modes in the model with
         homogeneous outer layers. In the inhomogeneous case,  
         non-successive overstable radial orders exist.
}
\endfigure

To see where the loss of the $\delta$ Sct modes in inhomogeneous
envelope models comes from, we look at the work integral for the $k=9$
dipole mode with a period of 3\,560~s of model L. In the homogeneous
envelope model this mode is overstable but it is damped in the
inhomogeneous one.  Figure~11 displays in the lower panel the cumulative
work; the solution for the homogeneous case is delineated with the
full and the inhomogeneous one with the dashed line. Both work
integral curves are independently normalized to unity at the
surface. Obviously, the hydrogen partial ionization zone contributes
to driving in the homogeneous but damps in the inhomogeneous envelope,
and it is the action of this region which decides over stability or
overstability.  We attribute the swapping of the work contribution of
the H-ionization zone to the influence of the different envelope
structures. Notice that the $\kappa_{\rm T}$ runs are very close in
both cases. Therefore, it is the phase relation between density and
pressure perturbation at the H-ionization zone which cause the
difference. Notice once more that that He drainage extends only to
$\log P \approx 5$.

\beginfigure{11}

\epsfxsize = 8.3 cm
\epsfbox{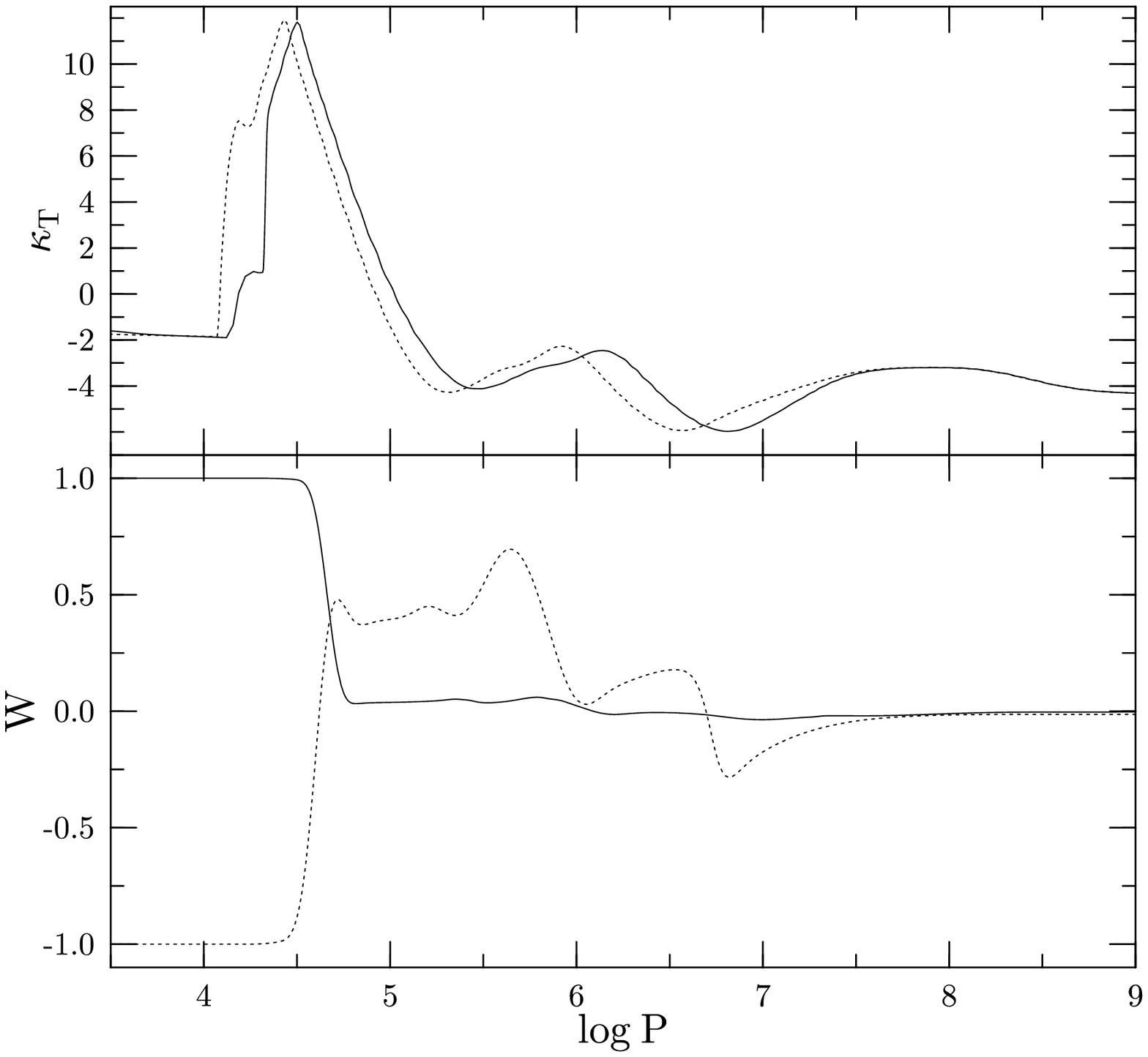}

\caption{{\bf Figure 11.} 
        Effect of He drainage in the outermost part of the 
        envelope on the work integral of $\ell=1, k=9$ mode.
        The full line depicts the solution on the homogeneous 
        envelope. The dashed line stands for the one on
	the chemically inhomogeneous envelope.
        }
\endfigure

\subsection{ Dependence on spherical degree}

Observed frequency spectra of roAp stars hint at the possibility that
not only dipole modes are involved in their light variability.  In
some cases~--~such as HR~1217, HD~119027, HD~203932~--~the frequency
spacings are not even enough to be explainable with $\ell = 1$ modes
only (Kurtz 1995). Also from the theoretical side, assuming now that a
classical $\kappa$ mechanism drives the modes, we expect not only
$\ell = 1$ modes to be excited.

\beginfigure{12}

\epsfxsize = 8.3 cm
\epsfbox{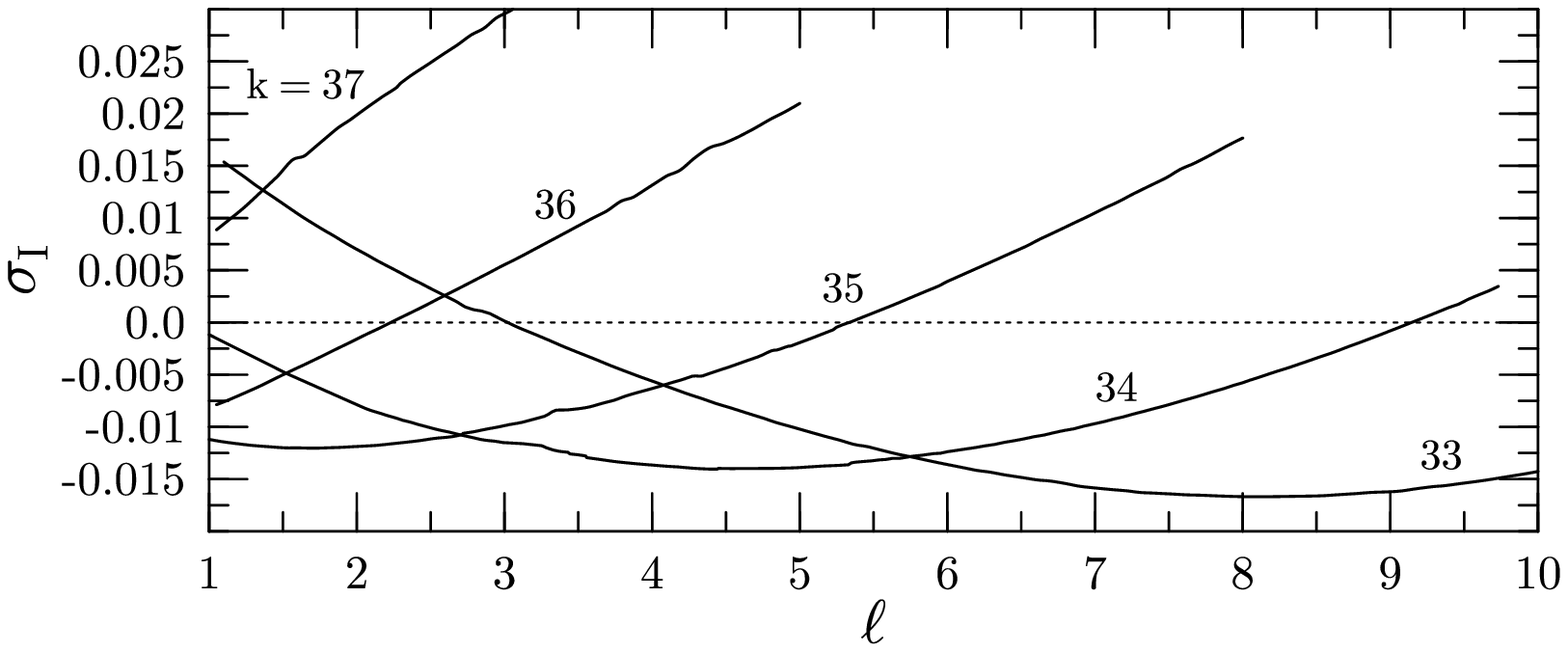}

\caption{{\bf Figure 12.} 
         Imaginary parts of high-order \p~modes of model L as a 
         function of spherical degree. 
        }
\endfigure

Figure~12 shows results for the inhomogeneous model L with weak
temperature inversion. We plotted directly the imaginary parts of the
eigenfrequencies for radial orders 33 to 37. The highest order which
becomes overstable at $\ell = 1$ is 36. Lower orders tend to have
their maximum driving at higher degrees.  
From the asymptotic relation $\nu_{k\ell} \propto \nu_0 ( k + \ell / 2 +
\varepsilon)$ we deduce that the maximum driving occurs always at similar
period, irrespective of the spherical degree.  Our computations showed that a
mix of different spherical degrees should contribute to the roAp variability.
As it is monitored mainly photometrically at the moment, spherical degrees
higher than 3 are not very relevant. Adding up all the low-$\ell$ (0 to 3)
modes and the different radial orders, we found that usually less than 12
modes (in terms of the principal quantum numbers $k$ and $\ell$) are
simultaneously excited.

\subsection{Node distribution in the atmospheres}

\beginfigure{13}

\epsfxsize = 8.3 cm
\epsfbox{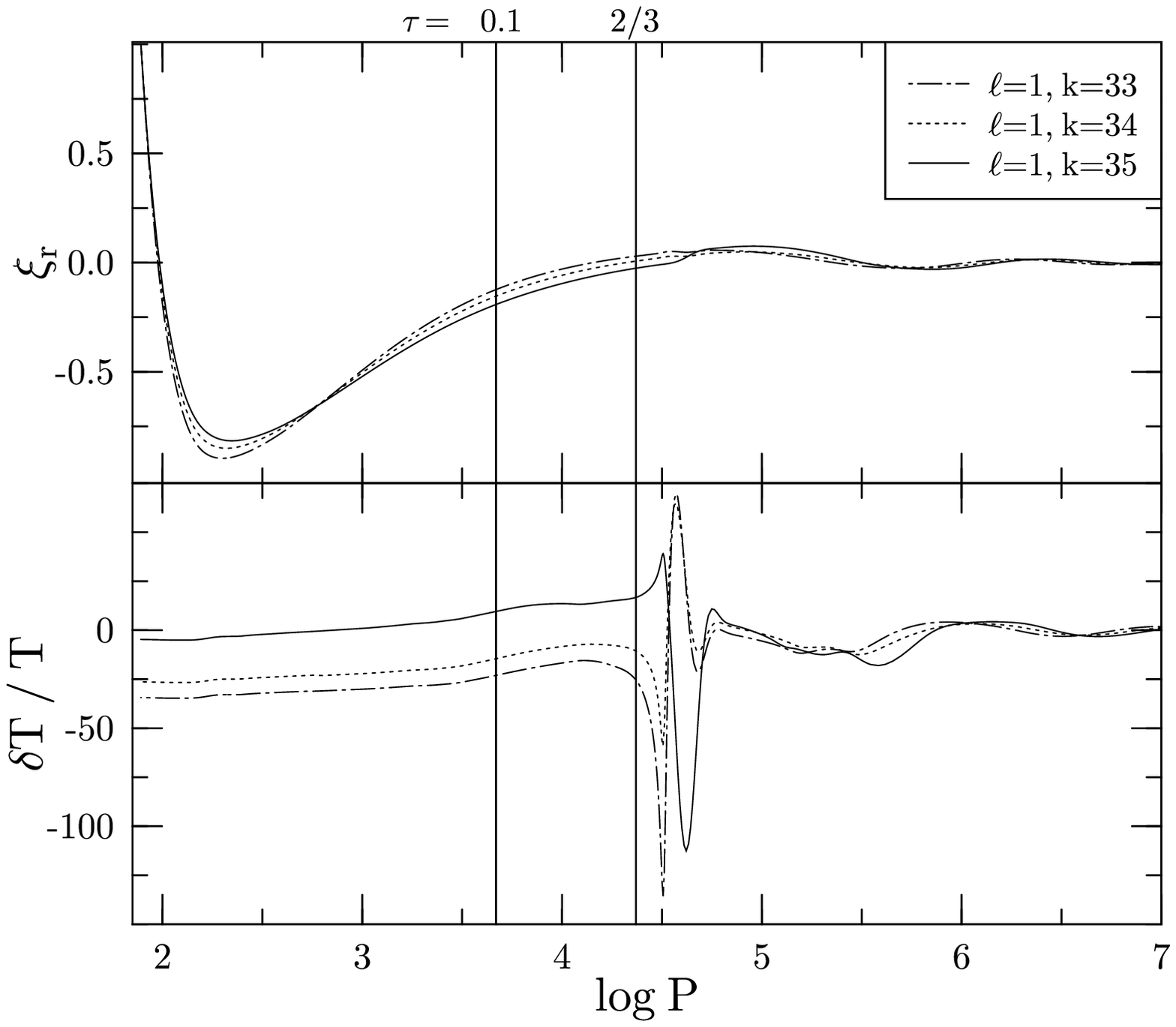}

\caption{{\bf Figure 13.} 
         Spatial behavior of eigenfunction in the outermost 
         layers for the 3 overstable high-order dipole modes 
         of model H.
        }
\endfigure
For recent spectroscopic observations of $\alpha$~Cir, phase shifts in the
temporal behavior of absorption-line features which probe different depths in
the atmosphere suggested that a node is lying in the optically thin
superficial layers (Baldry et al. 1998).  We analyzed our eigensolutions in
this respect. Figure~13 shows the overstable dipole modes of model H.  The
upper panel depicts displacement eigenfunctions. They do not shift
significantly for successive radial orders. In all three cases displayed, a
node lies at very low optical depths and a second one close to $\tau = 2/3$.
The situation is different for the temperature perturbation for which the real
part of the eigenfunction is shown in the lower panel of Fig.~13. The $k=35$
mode shows a node at small optical depths, whereas the next two lower-order
modes have no node at $\tau < 2/3$. The reason for the temperature
perturbation to have a stronger variation than the displacement function is
the very superficial H/He ionization zone. This region causes, due to a strong
entropy perturbation, an associated strong temperature variation.

We emphasize that our treatment of the outer regions, in particular of
the optically thin regions are far from realistic. Therefore, the
results in Fig.~13 should at best be taken as indications that indeed
nodes can be expected in the atmospheres of roAp stars. Quantitative
aspects will certainly change when studies concerning this issue are
performed with the correct perturbed transport equations. Only then
can we expect to exploit this property of roAp stars for atmospheric
diagnosis.

\section{Discussion}

Nonadiabatic stability analyses applied to evolutionary models with
homogeneous Pop I composition revealed that the $\kappa$ mechanism of H/He
ionization can destabilize short-period roAp-type modes.  We found the
short-period instabilities to be restricted to the 1.5 and 1.6 $\msol$
sequences. The rough location of the blue edge for the high-order roAp
oscillations lies, according to our preliminary computations, at about $\log
\teff = 3.86$.  Both, main-sequence and early subgiant models can become roAp
pulsators. In contrast to a recent mapping of roAp stars onto the HR plane
(North et al. 1997), we find our overstable models to have somewhat lower
masses and to have lower effective temperatures than what the North et
al.(1997) data imply.  On the other hand, we expect the blue edge to shift to higher temperatures
if the more helium (than what we assumed in our computations) is depleted from
the most superficial stellar layers. In this case the H driving in the H/HeI
partial ionization zone should get more efficient. This has to be proven,
however, with explicit computations in the future.

In contrast to $\delta$ Sct stars, H/He ionization dominates the driving of
the roAp modes.  Since the H/He ionization zone lies at lower temperatures
than the He$^+$ one, we expect the blue edge of the roAp instability domain to
be {\it somewhat\/} cooler than that of $\delta$ Sct stars (cf. Fig.~1).

In all cases, the number of overstable oscillation modes we counted in
our computations was small. Not more than 5 overstable modes of a
given spherical degree were encountered. This is still larger than
what is observed in some roAp stars so that mode selection remains a
problem to be addressed. In contrast to $\delta$ Sct stars, however,
the number of overstable modes is considerably smaller in roAp
models. From Fig.~5 we see that at the base of the $\kappa_{\rm T}$
gradient induced by H/He partial ionization the thermal scale
levels off at about 1\,000 s. Only periods with shorter periods can
be potentially excited in this region. As the overtones are already
high, the radiative dissipation is also high. Therefore, it seems
that only for a few modes with high radial order can H/He driving
overcome the radiative dissipation anymore. 

A necessary prerequisite for short-period oscillations to be excited was
the hypothesis of a temperature inversion in the inner atmosphere
of the models. The temperature inversion which raises the superficial
critical frequency sufficiently to reflect roAp-type modes need not
be very high. Depending on the steepness, one to three thousand K are
enough. Another effect of the steepness of the inversion is that it
influences the number of excited modes.

The evolutionary stage of roAp stars can be anywhere between the main
sequence and the beginning of the subgiant stage~--~observed
frequencies do allow for this bandwidth of dating. In the most
advanced phases of evolution the core is no longer convective.
Therefore, no magnetic dynamo could be operative anymore; most of the
star~--~except for the narrow ionization regions~--~is radiative. If
the cyclic secular frequency variations of roAp stars were interpreted
to be due to some magnetic activity cycle then the ohmic decay time of
the stellar magnetic field would limit the maximum age of such roAp
stars. At present, the origin of the frequency variations as observed
by now in HD~134214 and HR~3831 (Kurtz et al. 1994, Kreidl et
al. 1994, Kurtz et al. 1997) is far from clear.

From our computations we cannot unambiguously attribute an
evolutionary stage to the range of periods of overstable oscillation
modes. From Fig.~7 we see a tendency of longer periods to be excited
at later evolutionary stages in the $1.5 \msol$ models. This finding
could, however, not be reproduced in the $1.6 \msol$ models.

Introducing an atmospheric temperature inversion in the models
eventually produced the long elusive overstable roAp-type
modes. However, in the same models we found roAp and $\delta$ Sct
modes excited~--~this contradicted observational evidence. Therefore,
we argued for a He drainage in these superficial regions where magnetic
pressure dominates over gas pressure. When assuming a kilo-Gauss
dipole field we found a mass fraction of only a few times $10^{-9}
\mast$ to be magnetically dominated and hence drained from He. In such
models we got rid of the $\delta$ Sct modes but could retain the roAp ones. As
only the the H/He ionization was deficient of He, it is evident that H
ionization dominates the excitation of roAp modes (see also the work integrals
in Fig.~8). Notice that the He$^+$ partial ionization is {\it not} modified in
these models, but still we got rid of the $\delta$ Sct-type modes. This
contradicts previous reasoning (Cox et al. 1979). Our computations of
inhomogeneous models showed that for $\delta$ Sct-type modes the H/He
ionization zone has a damping influence which even over-compensates the
driving action of the He$^+$ region. Furthermore, observations support our
picture which emphasizes the r\^ole of the magnetic field for the presence of
high-frequency (roAp-like) oscillation modes which are not found in $\delta$
Sct stars such as in FG Vir (Breger et al. 1996).

In some of the inhomogeneous models we found excited non-successive overtones.
Such a behavior is reminiscent of the frequency spectrum as found for
example in HD~217522 (Kurtz 1995).

A major problem with whole stability analysis as presented here remains: The
driving region of the roAp modes reaches into the region where the magnetic
pressure dominates the star, therefore we must expect a non-vanishing
influence of the perturbed magnetic field. In other words, magneto-acoustic
modes might be excited which carry away energy (Roberts \& Soward 1983,
Dziembowski \& Goode 1996). If such an interaction is efficient enough it can
damp out the pure acoustic modes of which we were talking here. However, at
the moment we are not in a position to do such a nonadiabatic analysis
quantitatively. Dziembowski \& Goode (1996) presented estimates of such an
interaction in the adiabatic limit.  From their results we expect that only
the very strongest excited modes might survive in the correct hydro-magnetic
treatment.

Martinez (1996) discussed an unusual, possible roAp
star~--~HD~75425~--~which oscillates with 30-min period.  This period
is significantly longer than what is found for roAp stars and somewhat
short for $\delta$ Sct stars. Referring to Fig.~7 we find that such
long periods could indeed be excited.  For them the action of the
He$^{+}$ is important, however. If we interpret HD~75425 in our
framework, it should either be rather young so that sedimentation was
not yet efficient enough or the stabilisation of the flow by the
magnetic field is not sufficiently strong. The latter alternative
would require a magnetic field which is weaker than what is usually
found in other roAp stars. It is, however, not yet clear {\it why\/}
no high-order modes are present in HD~75425. Either they have not yet
been seen and they are actually hidden (at even lower amplitude) in
the long-period data or the hypothesized temperature inversion is not
strong enough to trap the high-order \p~modes.

It appears highly rewarding to search for weakly magnetic A-type stars to see
if it is possible at all to have both period domains (cf. Figs.~6 and 7)
excited in any such stars. To extend our speculations further in this
direction we mention Am~--~Fm stars (Kurtz et al. 1995, Kurtz 1998). These
mildly peculiar cool main-sequence stars seem to be always binaries.
Sedimentation of elements due to suppressed rotational velocities might
eventually suppress $\delta$ Sct pulsations. Since the chemical peculiarity is
not associated with a magnetic field these stars cannot build up a
sufficiently high critical frequency to support roAp-type modes. Therefore,
we do not expect Am~--~Fm stars to be seen as short-period pulsators; on the
other hand, if He drainage is not strong yet (i.e. if they are young enough)
they might well be $\delta$ Sct type pulsators (cf. Cox et al. 1979).

\beginfigure{14}

\epsfxsize = 8.3 cm
\epsfbox {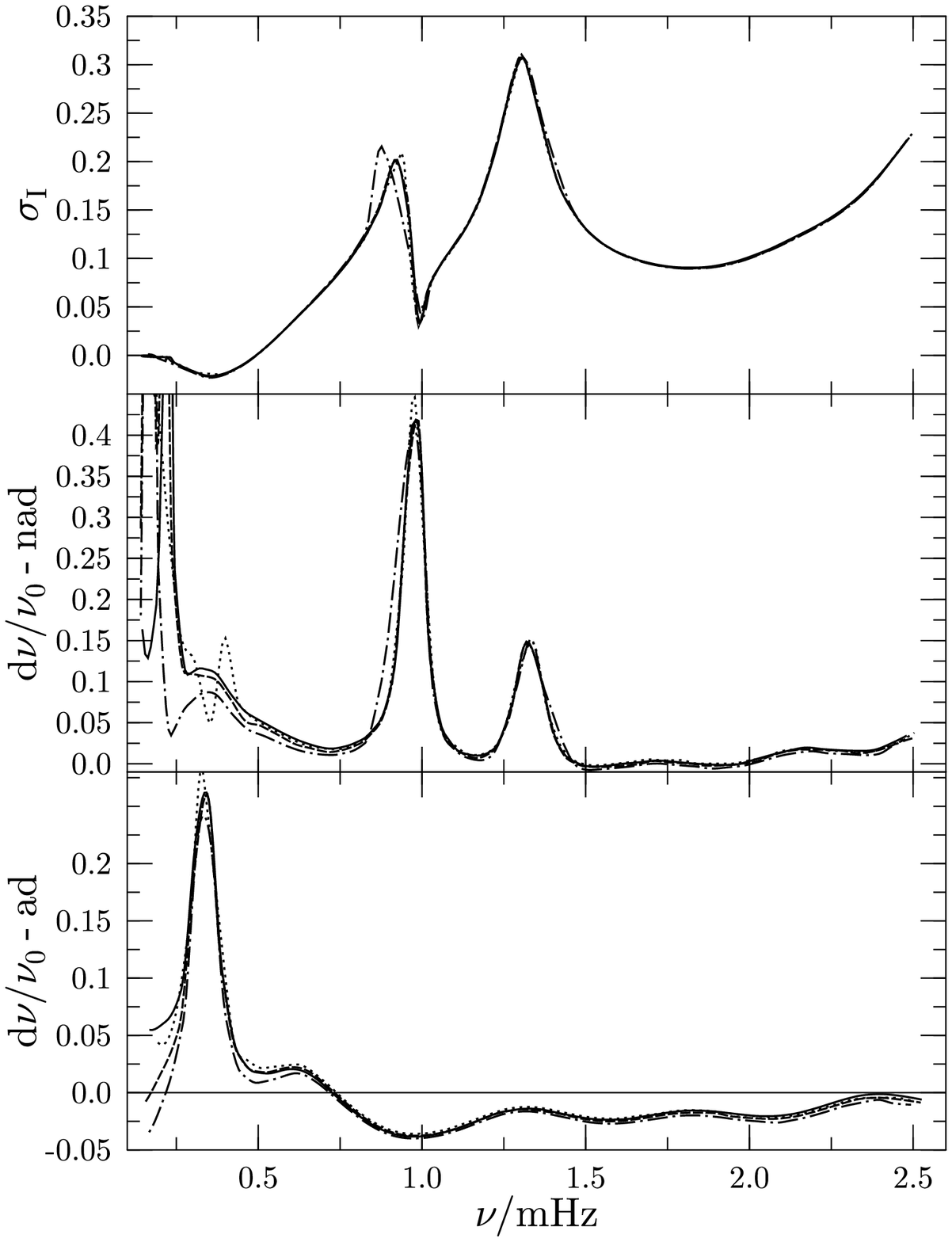}


\caption{{\bf Figure 14.} 
         Frequency behavior for model B. The top diagram shows the
         imaginary parts of the nonadiabatic eigenvalues.
         The values for degrees $1 \le \ell \le 4$ are plotted
         on top of each other. The middle panel shows the fractional
         deviation of the nonadiabatic frequency spacing from the
	 asymptotic value. The lowest panel shows the same deviation
         computed from the adiabatic frequencies. The line coding
         is as following: $\ell = 1$: full line; $\ell = 2$: dotted
         line; $\ell = 3$: dashed line; $\ell = 4$: dash-dotted line. 
        }
\endfigure

As mentioned before, we were able to enhance the critical frequency at
the stellar surface by ad hoc postulating a temperature inversion in
the optically thin regions of the stellar atmosphere.  Physically,
this is equivalent to assuming a chromosphere in these stars. The
spectral types of the roAp stars are just at the border where
chromospheric evidence starts to be observed. According to prevailing
theoretical ideas, the roAp stars provide already the necessary
ingredients for chromospheric heating: a superficial convection zone
and strong magnetic fields which might provide the necessary acoustic
flux and possible magneto-acoustic waves, respectively. Shore et
al. (1987) found, however, no signs of chromospheric emission in roAp
stars.  For further progress it would be highly rewarding to find
observational evidence for or against the temperature-inversion 
hypothesis.

The appendix discussed the approach of nonadiabatic frequencies to what is
expected from adiabatic asymptotics. We found that caution is required when
using high-frequency oscillation data to infer stellar parameters. Even if the
excitation rates of high-order modes are low, the deviations of frequency
spacings from the adiabatically predicted ones can be up to about $8\%$ (see
Fig.~15). Furthermore, this behavior is expressedly non-monotonic so that an
easy correction cannot be applied. The magnitude of nonadiabatic frequency
shifts in roAp stars is not negligible compared with the magnetic-field
induced frequency perturbations (Dziembowski \& Goode 1996). Therefore,
asteroseismology of roAp stars is quite a challenge for pulsation theory. To
finish, we feel that a warning of simple application of adiabatic asymptotics
to high-order modes is also in place for solar seismology.

\section*{Acknowledgments}

Financial support by the Swiss National Science Foundation through a PROFIL2
fellowship (A.G.) is gratefully acknowledged. For part of this work, H. Saio
was supported by the Swiss National Science Foundation and the Japan Society
for the Promotion of Science. Enlightening discussions with and constructive
refereeing of the manuscript by D.~Kurtz as well as the initial motivation by
H.~Shibahashi were of great help. W. L\"offler kindly provided most of the
stellar models used for the analyses.

\section*{Appendix}

\beginfigure{15}

\epsfxsize = 8.3 cm
\epsfbox {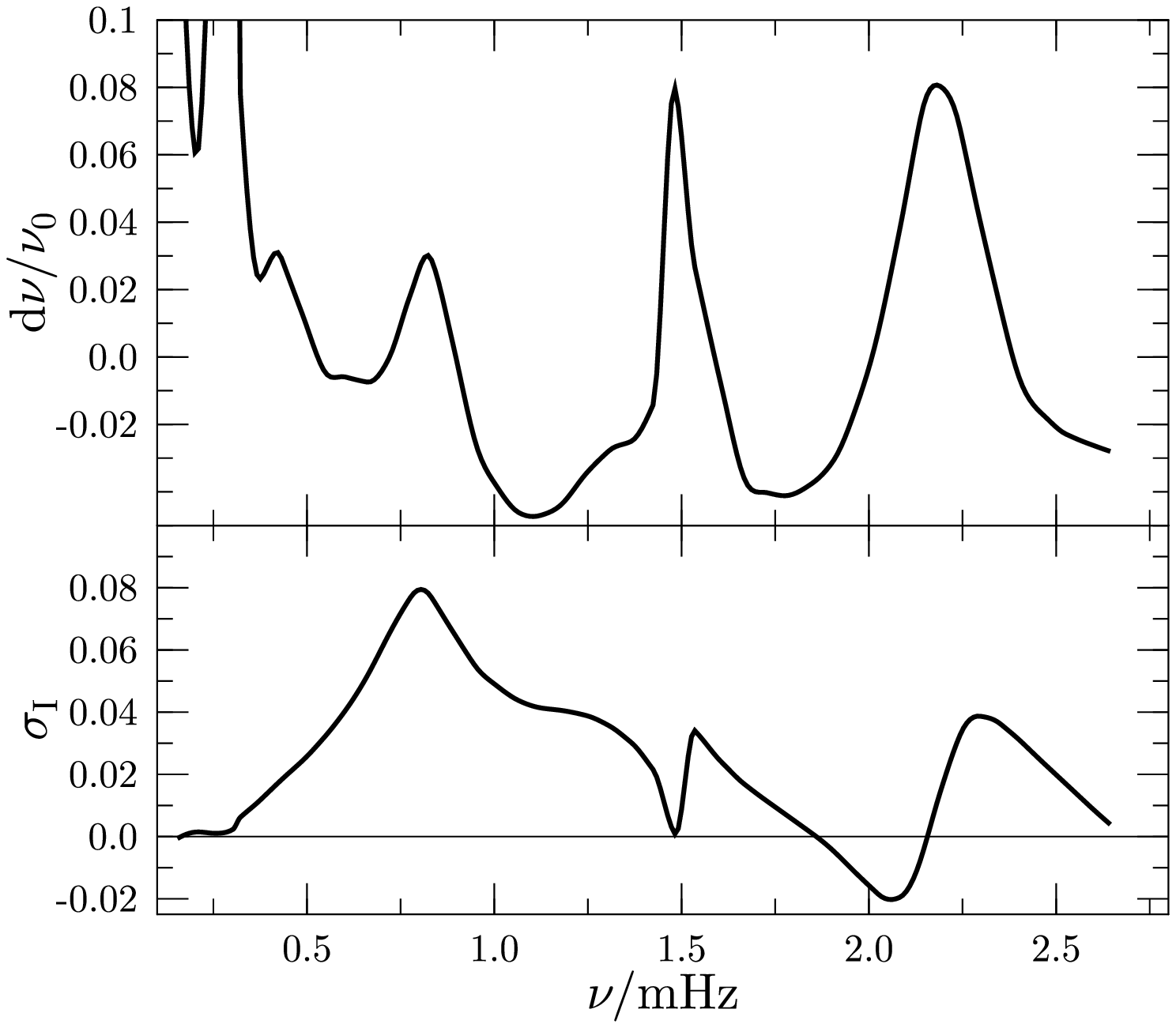}


\caption{{\bf Figure 15.} 
         Relative deviation of the frequency spacing from the
         asymptotic value for model K. 
        }
\endfigure

We discuss concisely the effects of nonadiabaticity on the
oscillation-frequency spectrum and compare it with adiabatic
results. This is mainly of interest for those aiming at using roAp
frequency data for seismological deductions.

For high radial-order modes of low degree, the frequency separation
of equal-degree modes is usually rather close to  
$$
{1 \over \nu_0} = 2 \int_0^{R_\ast} {\diff r \over  c_{\rm ad}}
\eqno(A1)
$$
with the adiabatic sound speed given by $c_{\rm ad} \equiv
\sqrt{\Gamma_1 P / \rho}$ (cf. Sect.~3.4). The value of $\nu_0$ 
corresponds to the frequency of an acoustic wave to travel across a
star.

In the following we investigate how quickly the frequency separation
approaches the asymptotic limit which is frequently referred to for
asteroseismic purposes.  To quantify the relative deviation of the
actually computed frequency spacing from the expected one ($\nu_0$),
we introduce
$$
{{\rm d} \nu \over \nu_0} = 1 - {\nu_{k+1} - \nu_k \over \nu_0}.
\eqno(A2)
$$

Figure~14 shows extensive results for model B for which we computed
oscillation modes with spherical degree 1 (full line), 2 (dotted
line), 3 (dashed line), and 4 (dash-dotted line) for many radial
orders.  Computations were performed adiabatically as well as
non-adiabatically. The bottom panel of Fig.~14 displays the relative
deviation ${\rm d} \nu / \nu_0$ of the adiabatic frequency spacing as
a function of the oscillation frequency. Above about 1.3~mHz the
deviation from the asymptotic value (51.46~$\mu$\,Hz) drops below 2 \%
for all degrees.  In the range between 2.25 and 2.5~mHz, where many
roAp modes are observed, the adiabatic mode separation agrees with the
asymptotic value on the level of 0.1 to 1~\%, depending on the
degree. This frequency range corresponds to radial orders exceeding
40.

To see if the convergence properties of the nonadiabatic
eigenfrequencies are comparable with the adiabatic ones we plot the
relative deviation from the asymptotic value (adiabatically computed)
in the middle panel of Fig.~14. The most obvious discrepancy is seen
around 1 and 1.3 mHz. There we find a relative deviation of the
spacing of 40~\% and 10~\%, respectively. These two features are clearly
associated with the local maxima in the damping rates $\sigi$ which
are shown in the top panel of Fig.~14. In the frequency range between
1.5 to 2 mHz, the frequency separation relaxes to within a few percent
of the asymptotic one.  For higher frequencies, however, they start to
diverge again in accordance with the rise of the damping rates.

What relevance does this behavior have for roAp modes? After all, the
modes which we used for the above demonstration were all damped.  We
consider now the relative deviation ${\rm d} \nu / \nu_0$ of the
nonadiabatic oscillation frequency separation from the the adiabatic
asymptotic value for $\ell = 1$ of model K. This model has 4
overstable modes around 500~s period. The imaginary parts are rather
small but still the frequency spacing reacts sensitively to the
$\sigi$ variation. Figure~15 shows that at 1.5~mHz~--~at the position
of a marginally overstable oscillation mode for which the damping rate
is close to zero~--~the frequency spacing deviates by about 8 \% from
the adiabatically expected one. In the overstable period domain, the
frequency spacings change from 3~\% longer to 8~\% shorter than the
predicted asymptotic frequency spacing. For seismological inferences
such differences can be relevant.

To finish, we emphasize that, if high accuracy is crucial, nonadiabaticity
effects are not negligible in deriving stellar parameters from 
high-frequency oscillation data. This conclusion might not only
apply to roAp stars but also to the high-frequency domain of
solar oscillations.

\section*{References}

\bye